\documentclass[10pt,a4paper]{article}
\usepackage{geometry}
\usepackage[utf8]{inputenc}

\usepackage[normalem]{ulem}
\usepackage{multicol}
\usepackage{diagbox}
\usepackage{caption}
\usepackage{subcaption}
\usepackage{jcappub}
\usepackage{bm}
\usepackage{epsfig}
\usepackage{bbold}
\usepackage{graphicx}
\usepackage{amsmath}
\usepackage{amssymb}
\usepackage{amsbsy}
\usepackage{color}
\usepackage{slashed}
\usepackage{afterpage}
\usepackage{psfrag}
\usepackage[noabbrev]{cleveref}
\usepackage{dsfont}
\usepackage{enumerate}
\usepackage[shortlabels]{enumitem}
\usepackage[utf8]{inputenc}
\usepackage{amsmath}
\usepackage[dvipsnames]{xcolor}
\usepackage{graphicx}
\usepackage{braket}
\usepackage[most]{tcolorbox}
\usepackage{empheq}
\usepackage{physics}
\usepackage{hyperref}

\newcommand{\excluir}[1]{}

\def\bt{\bar t}
\def\bz{\bar \theta}

\title{\begin{center}
        Cuspidal Singularities in Collapsing Domain Walls
\end{center}}

\vfill
\author[a,b,c]{Jose J. Blanco-Pillado,}
\author[b,c,d]{Matthew Elley,}
\author[e]{Francesc Ferrer,}
\author[b,c,f]{Alberto García Martín-Caro,}
\author[g]{Daniel Jiménez-Aguilar,}
\author[h]{Oriol Pujolàs,}
\author{and}
\author[h]{Juan S. Valbuena-Bermúdez}

\affiliation[a]{IKERBASQUE, Basque Foundation for Science, 48011, Bilbao, Spain}
\affiliation[b]{EHU Quantum Center, University of the Basque Country, UPV/EHU}
\affiliation[c]{Department of Physics, University of the Basque
Country, UPV/EHU, 48080, Bilbao, Spain}
\affiliation[d]{CEICO, Institute of Physics of the Czech Academy of Sciences, Na Slovance 1999/2, 182 00, Prague 8, Czechia}
\affiliation[e]{Department of Physics and McDonnell Center for the Space Sciences,
Washington University, Saint Louis, MO 63130, USA}
\affiliation[f]{Instituto de Física, Computación e Ciencias Aeroespaciais (IFCAE), Universidade de Vigo. 32004 Ourense, Spain}
\affiliation[g]{Institute of Cosmology, Department of Physics and Astronomy, Tufts University, Medford, MA 02155, USA}
\affiliation[h]{Institut de Física d'Altes Energies (IFAE) and Barcelona Institute of Science and Technology (BIST), Campus UAB, 08193 Bellaterra, Barcelona, Spain}

\emailAdd{josejuan.blanco@ehu.eus}
\emailAdd{elley@fzu.cz}
\emailAdd{ferrer@wustl.edu}
\emailAdd{alberto.garcia.martin-caro@uvigo.gal}
\emailAdd{Daniel.Jimenez\_Aguilar@tufts.edu}
\emailAdd{pujolas@ifae.es}
\emailAdd{jvalbuena@ifae.es}

\makeatletter
\gdef\@fpheader{}
\makeatother

\abstract{Domain wall networks have attracted renewed interest, particularly in relation to the dynamics of network collapse. Accurately describing this process is challenging and typically requires large scale numerical simulations. Here we adopt a complementary approach by studying the collapse of individual closed domain walls, extending previous thin wall analyses and comparing them with adaptive mesh refinement field theory simulations. Firstly, we show that collapsing domain walls generically develop worldvolume singularities of two types: \textit{cuspidal edge} singularities, consisting of one dimensional singular edges that propagate along the wall surface at the speed of light for a finite time, and \textit{cuspidal vertex} singularities, which are spike like and instantaneous events where the wall moves momentarily at the speed of light. Both types of features arise generically from smooth initial conditions, and their formation and evolution follow the universal patterns of singularity theory. We show that these structures are captured both by the Nambu-Goto equations and by an eikonal like approximation valid in the relativistic regime. Furthermore, we demonstrate that the same singular structures are reproduced qualitatively in full field theory simulations, establishing that they are not artifacts of the thin wall approximation but robust features of realistic domain wall dynamics. Naturally, such focusing effects in the field theory simulations result in localized regions of high energy density. We briefly discuss possible phenomenological implications.}

\begin{document}

\maketitle


\section{Introduction}

Are there spontaneously broken discrete symmetries in nature? This question, relevant to any discrete symmetry, can be answered quite sharply. The spontaneous breaking of discrete symmetries leads to the formation of domain walls (DWs), which have a large impact on cosmology, see {\em e.g.}~\cite{Kibble:1976sj, Vilenkin:2000jqa, Vachaspati:2006zz} for reviews. Given that the DW energy is proportional to the area, once a DW network forms, it dilutes very slowly. Without a mechanism to eliminate them, the universe would be stuck in the DW domination phase, a highly inhomogeneous stage that is (difficult to study and) seemingly difficult to gracefully exit. This is the so-called DW problem~\cite{Zeldovich:1974uw, Vilenkin:1981zs}, implying that spontaneously broken exact discrete symmetries are basically ruled out. This conclusion is strongly sensitive to the exactness of the discrete symmetry. 
Any small explicit breaking (in the Lagrangian or in the initial conditions) provides a mechanism (pressure bias or population bias, respectively) by which the DW network annihilates after a certain time scale. Spontaneously broken approximate discrete symmetries (present in a variety of well-motivated models), then, are allowed~\cite{Sikivie:1982qv, Gelmini:1988sf}, and the cosmological impact of DWs becomes an asset to probe them. 

The main generic imprints of decaying DW networks on cosmology are primordial gravitational waves (GWs)~\cite{Gleiser:1998na, Hiramatsu:2010yz, Kawasaki:2011vv, Hiramatsu:2012sc, Krajewski:2016vbr, Krajewski:2017czs, Saikawa:2017hiv, Ferreira:2024eru, Notari:2025kqq, Cyr:2025nzf, Gruber:2024pqh, Dankovsky:2024zvs,Dankovsky:2024ipq} and primordial black holes (PBHs)~\cite{Garriga:2015fdk, Deng:2016vzb, Ferrer:2018uiu, Gelmini:2022nim, Dunsky:2024zdo, Gouttenoire:2025ofv} (see~\cite{Abac:2025saz} for a recent review). 
Additional motivation for the study of domain walls has emerged from recent pulsar timing array (PTA) observations, which have provided indications of the existence of a stochastic gravitational wave background~\cite{NANOGrav:2023gor, Reardon:2023gzh, EPTA:2023fyk, Xu:2023wog}. In this context, 
gravitational wave signals generated by domain wall networks have been identified as a viable class of scenarios that could contribute to the PTA signal~\cite{NANOGrav:2023hvm}. Thus, a detailed understanding of the microscopic dynamics underlying DW collapse and the associated gravitational wave emission is crucial for accurately characterizing the observational signatures of these cosmological sources.

Broadly, the evolution of a DW network follows two distinct regimes: scaling and annihilation. Soon after formation at the symmetry breaking transition, DWs reach a scaling regime characterized by having roughly one large DW per Hubble patch. In this regime, DWs basically interconnect with one another, and closed structures are rarely formed. The annihilation phase is radically different from the scaling epoch. The most studied annihilation mechanisms consist of pressure bias (a small explicit symmetry breaking in the Lagrangian) or a population bias (a small asymmetry in the initial condition). In both cases, the predominant motions in the network consist of the collisions of walls and the collapse of closed DWs of various shapes. 
The dominant GW signal arises from near the onset of annihilation. This is both because it is the least distant to us and because of the added violent motions. Recent numerical simulations that include the annihilation phase indeed confirm that the GW signal is around one order of magnitude larger~\cite{Kitajima:2023cek, Ferreira:2024eru, Cyr:2025nzf, Notari:2025kqq} than previous estimates~\cite{Hiramatsu:2010yz, Kawasaki:2011vv, Hiramatsu:2012sc, Saikawa:2017hiv}, though details differ among the various groups~\cite{Cyr:2025nzf, Notari:2025kqq, Babichev:2025stm, Dankovsky:2025pjg}. Interestingly, some groups find that the spectrum of GWs is not only characterized by a single scale~\cite{Cyr:2025nzf, Notari:2025kqq}.  

It should be emphasized that numerical simulations of domain wall networks are particularly challenging, as they must simultaneously resolve sufficiently extended scaling and annihilation regimes while the characteristic DW size continues to grow and the comoving wall width correspondingly decreases in the simulation. As a result, these types of cosmological numerical simulations necessarily require a very large dynamical range by their final stages. The other important yield from the annihilating DW network is primordial black holes. This is even more difficult to compute since it requires following the sparse remaining closed DW structures after most of the network breaks into isolated fragments, and of course, including the DW self-gravity.

In this work, we adopt a complementary strategy to investigate the collapse of domain wall networks. Rather than relying on a brute force simulation of the full network, we focus on the evolution of individual closed domain walls of generic shape and follow their contraction both in the thin wall approximation and in full field theory simulations. The most notable outcome of this analysis is that singularities develop on the worldvolume after a finite time, well before the walls reach their point of minimal size.

We show that the formation of these singularities is generic and unavoidable, arising even from smooth initial configurations, and that they are also reproduced in field theory simulations with sufficiently high resolution. In the language of singularity theory, the domain wall worldvolume develops {\it catastrophic} structures at which the wall locally reaches ultrarelativistic velocities. In the context of domain wall collapse, we refer to these configurations collectively as {\it cuspidal singularities}. Physically, they constitute the domain wall counterparts of cusps~\cite{Turok:1984cn} on cosmic strings, and, as in that setting, one may expect both substantial gravitational wave emission and copious radiation into heavy modes of the underlying microscopic field theory from which the soliton arises~\cite{Vachaspati:1984gt,Brandenberger:1986vj,Olum:1998ag}.

Moreover, in contrast to cusps on cosmic strings, the singularities that develop on domain walls are not always instantaneous. As a result, one may expect an even more significant production of hard modes during the evolution of the wall. Although in this work we do not attempt to establish a quantitative connection between these singularities and the resulting gravitational wave spectrum, it is natural to expect that their formation could lead to an enhancement of power in the high frequency region of the spectrum.

Let us briefly summarize the previous work on the collapse of closed DWs. Linearization of the Nambu-Goto (NG) equation for small perturbations around the background of a spherical collapsing DW concluded that perturbations grow along the collapse~\cite{Widrow:1989fe}. However, the formation of caustics and cuspidal singularities requires a nonlinear analysis of the NG equations.  Crucial progress was made in~\cite{Hoppe:1995sn, Eggers:2009zz, Singularities_rel}, showing that even though the NG equations are not integrable for DWs, they can be reduced to first order equations in an appropriate gauge. The equations take the form of a generalized eikonal equation, amenable to understanding the generic features of focusing and caustic formation. These works were also the first to report on the appearance of swallowtail-like singularities on the DW worldvolume (named after the distinctive shape akin to that of a Swallow's tail). These were shown to appear during the evolution of closed non-spherically symmetric DWs. In the present work, we will closely review Refs~\cite{Hoppe:1995sn, Eggers:2009zz, Singularities_rel}, extending some aspects of the domain wall dynamics in the ultrarelativistic limit of the NG approximation and discussing higher order singularities that appear in the course of the DW collapse. As mentioned above, our main contribution will be the discussion of how all these singularities persist in the field theory description. \\

The rest of this article is organized as follows. In Sec.~\ref{sec:NG} we review the thin-wall description of relativistic domain walls and the corresponding Nambu-Goto dynamics in $3+1$ dimensions. After introducing the general formalism, we study the simplest configuration that exhibits these singularities, the collapse of axially symmetric ellipsoidal walls, and show explicitly how worldvolume singularities arise in finite time. We then analyze the local structure of these events within the framework of singularity theory, identifying the associated cuspidal edge configurations and their relation to swallowtail type bifurcations. In Sec.~\ref{sec:EikoalApp} we turn to the ultrarelativistic regime and present the eikonal, or ray tracing, approximation, showing that it provides an accurate and geometrically transparent description of the collapse and correctly reproduces the singularity structure. In Sec.~\ref{genericity} we use this framework to address the genericity of the phenomenon, discussing separately the formation of the two types of singularities; the cuspidal edges and cuspidal vertices, and examining the resulting global evolution of collapsing closed walls. In Sec.~\ref{Sec:FT} we move beyond the thin wall approximation and study the problem in full field theory. After describing the numerical setup, we present adaptive mesh refinement (AMR) simulations showing that the same singular structures persist, albeit somewhat smoothed by the finite wall thickness. Finally, in Sec.~\ref{Sec:Conclusions} we summarize our results and discuss their possible implications for the dynamics and cosmological signatures of collapsing domain wall networks.\\

Simulation videos can be found in the following link \url{https://www.youtube.com/@DWCollapse}.

\section{Worldvolume Singularity Formation in the Nambu-Goto limit}
\label{sec:NG}

Worldvolume singularities of extended objects have been previously discussed in the literature, most notably in the context of the evolution of line-like objects such as cosmic strings. In this case, the integrability of the Nambu–Goto equations enables one to predict the formation of singular points along the string \cite{Turok:1984cn}. A defining feature of these singularities is that the string momentarily reaches the speed of light and develops a characteristic geometric structure, commonly referred to as a cusp \cite{Vilenkin:2000jqa,Blanco-Pillado:1998tyu}.

The evolution of extended objects can also be investigated in lower dimensional spacetimes, such as in 2+1 dimensions, to determine whether analogous singular behavior arises. This problem was examined in \cite{Blanco-Pillado:2025gzs}, where it was shown that higher order singularities generically emerge during the collapse of closed configurations (for related work, see \cite{Babichev:2026zca}). These singularities were classified within the framework of catastrophe theory as swallowtail singularities. A more detailed analysis reveals that such structures can be interpreted as arising from the formation of a special type of cusp singularity — {\it a bifurcation cusp} — followed by the emergence of a pair of cusps that propagate away from the original bifurcation point. Importantly, it was demonstrated in \cite{Blanco-Pillado:2025gzs} that these features are not artifacts of the Nambu–Goto description, but are reproduced with high accuracy in full field theory simulations of the corresponding line like solitonic configurations in $2+1$ dimensions.

In the present work, we investigate the formation of analogous singularities in the evolution of domain walls in 3+1 dimensions. As a first step, we analyze their possible emergence within the Nambu–Goto framework. In this section, we closely follow the formulation of domain wall dynamics developed in \cite{Hoppe:1995sn,Eggers:2009zz}.\\

We begin by describing the evolution of domain walls in flat spacetime within the framework of the Nambu–Goto action. To this end, we represent the wall by its worldvolume, a three dimensional Lorentzian hypersurface embedded in the ambient spacetime, specified by the mapping
\begin{equation}
X^\mu(\zeta^\alpha)~,
\end{equation} 
parametrized by worldvolume coordinates $\zeta^\alpha=(\tau,\sigma^a)$ with $\sigma^a=(\sigma_1,\sigma_2)$.
The Nambu-Goto action for walls in $3+1$ dimensions is given by
\begin{equation}
S=-\sigma_{DW}\int d^{3}\zeta\sqrt{-\gamma}\,,
\label{eq:NG action}
\end{equation} 
where $\sigma_{DW}$ is the energy per unit area of the wall and $\gamma$ is the determinant of the worldvolume metric, which  is given by
\begin{equation}
\gamma_{\alpha\beta}=\eta_{\mu\nu}\partial_{\alpha}X^{\mu}\partial_{\beta}X^{\nu}\,.
\label{eq:worldvolume metric}
\end{equation}

The equations of motion arising from the action (\ref{eq:NG action}) are
\begin{equation}
\partial_{\alpha}\left(\sqrt{-\gamma}\;\gamma^{\alpha\beta}\,\partial_{\beta}X^{\mu}\right)=0\,.
\label{eq:eom}
\end{equation} 

Since the Nambu-Goto action is invariant under reparametrizations of the worldvolume, we can freely constrain the six independent components of $\gamma_{\alpha\beta}$ (i.e, fix the gauge). In the case of strings, one chooses $\gamma_{\alpha\beta}$ to be conformally flat (this is always possible, as the induced metric is two-dimensional), with the additional choice
\begin{equation}
X^{0}=\zeta^{0}=t\,.
\label{eq:X0}
\end{equation} 
We also use this last condition in our analysis. However, for the problem at hand, the imposition of conformal flatness is not permissible, as it generically leads to an overconstrained system. Instead, we choose $\gamma_{\alpha\beta}$ to be block-diagonal with respect to $t$ and $\sigma^a$ directions, that is, we assume $\gamma_{0a}=0$. This leads to the constraint  
\begin{equation}
\dot{\mathbf{x}}\cdot\partial_{a}\mathbf{x}=0 \,, \label{eq:gamma 01 constraint} 
\end{equation}
where $\mathbf{x}=(x,y,z)=X^i$ and dot denotes derivative with respect to $t$. This constraint admits a straightforward physical interpretation: the physical motion of the domain wall is everywhere orthogonal to its tangent vectors.
The temporal and spatial blocks are 
\begin{equation}
    \gamma_{00}=-1+\dot{\mathbf{x}}^2 \qquad {\rm and}\qquad g_{ab} \equiv
    \gamma_{ab}=\partial_{a}\mathbf{x}\cdot\partial_{b}\mathbf{x}~,
\end{equation}
where for clarity we introduce the notation $g_{ab}$ for the spatial part.

Next, introducing \eqref{eq:X0} in the temporal component of \eqref{eq:eom}, one realizes that this equation takes the form of a local conservation law
\begin{equation}\label{cons}
    \partial_0 \left(\sqrt{-\gamma^{00}\;g}\right)=0 \qquad\Rightarrow \qquad
    -\gamma^{00}\;g=\rho^2(\sigma^a)
\end{equation}
with $g=\det(g_{ab})$ and $\rho$ is a $t-$independent function on the DW that can be written in terms of the initial conditions
\begin{equation}
\rho^2(\sigma^a) = \frac{g_0}{1-\dot{\mathbf{x}}_0^2}~.\label{eq:rho}
\end{equation}

This function admits a clear physical interpretation as the local energy density of the domain wall system, corresponding to the wall area weighted by the local relativistic Lorentz factor.

Eq. \eqref{cons} then implies the following first-order equation for the evolution of the spatial embedding of the wall:
\begin{equation}\label{const2}
    \dot{\mathbf{x}}^2+\frac{g}{\rho^2}=1~.
\end{equation}

This condition is reminiscent of one of the constraints for cosmic strings in conformal gauge, $\dot{\mathbf{x}}^2+{\mathbf{x}'}^2=1$, upon replacing ${\mathbf{x}'}^2$ by the determinant of the spatial metric, $g$, and using the residual reparametrization invariance to fix the spatial parameter along the string such that $\rho=1$ in that case.

Examining this constraint in the case of strings, one finds that points moving at the speed of light lead to singularities on the worldsheet, since  ${\mathbf{x}'}^2=0$. These correspond to the well known cusp configurations on strings~\cite{Turok:1984cn}, at which the induced metric becomes degenerate. The generic form of the string at these singularities can be described by an expansion of the form $y\propto x^{2/3}$ in an appropriate coordinate system \cite{Vilenkin:2000jqa}. In other words, the configuration exhibits the profile of a cusp, which motivates the terminology\footnote{In $2+1$ dimensions, alternative types of cusps may arise, characterized by scalings of the form 
$y\propto x^{3/4}$, commonly referred to as bifurcation cusps \cite{Blanco-Pillado:2025gzs}. Although such structures can also occur in strings in $3+1$ dimensions, they would require highly fine tuned configurations. Additional similar examples of finely tuned configurations exhibiting other types of cosmic string cusps in $3+1$ have been recently discussed in \cite{Drew:2025txu}.}.

One way to develop intuition for the emergence of these singularities in $2+1$ dimensions is to consider the formation of caustics during the collapse of closed configurations driven by their own tension. This perspective strongly suggests that an analogous phenomenon should arise in the DW case in $3+1$ dimensions.

A more direct way to see this is to examine \eqref{eq:gamma 01 constraint}, which requires that $\dot{\mathbf{x}}$  be proportional to the unit normal vector to the surface, $\mathbf{n}$. In this formulation, the Nambu–Goto equation can be understood as a first-order “normal flow”, or eikonal-like, equation where

\begin{align}\label{flow}
    \dot{\mathbf{x}} =& \,V \; \mathbf{n}\,,\qquad {\rm with}\\[3mm]  
    V\equiv&\sqrt{1-\left(1-\dot{\mathbf{x}}_0^2\right)\;\frac{g}{g_0}}~.\label{V}
\end{align}
For DW collapse problems it is clear that $\mathbf{n}$ is inward pointing so one is inclined to believe that singularities similar to the $2+1$ case would appear now as well.

Furthermore, this form of the equation of motion makes explicit that at degenerate singular points, $g=0$, the motion of the wall acquires the speed of light, namely  $|\dot{\mathbf{x}}|=1$. These are the analogue of the string cusps, although as we will see shortly these singularities come about along lines of the worldvolume in the case of domain walls, so we call them {\it cuspidal edges} to make this point more clear. In the following we will demonstrate that this type of singularities forms dynamically in the case of domain wall collapse. 

Moreover, it is also clear that there are various types of singularities, depending on the number of vanishing eigenvalues of $g_{ab}$ where $g=0$. The most common situation will be that one eigenvalue ({\it cuspidal edges}) vanishes, but examples of double degeneracy can be realized ({\it cuspidal vertex} ).
Let us now review the simplest class of domain wall configurations in which these singularities arise. The most elementary example of domain wall collapse is provided by a spherically symmetric configuration. In this case, the evolution unavoidably produces a singular point at the center of the wall, where all wall elements converge inward at ultra-relativistic velocities, resembling the singular structures analyzed in this work. Nevertheless, such a setup is highly idealized and is not expected to be representative of generic collapse, since the resulting singularity is largely a consequence of the fine tuned symmetry of the initial conditions.

A more representative configuration is provided by an axially symmetric ellipsoidal domain wall. As we will demonstrate, this case already exhibits nontrivial dynamics and gives rise to phenomenology that can be extrapolated to more general situations. This is the scenario we analyze next.

\subsection{Axially Symmetric Ellipsoid}\label{Sec:Ellipsoid}

For closed walls topologically equivalent to the sphere it is convenient to use the spherical angles as spatial parameters  $\sigma^a=(\theta,\varphi)$ on the worldvolume.
In that case, a general axially symmetric wall reduces to
\begin{equation}
\mathbf{x}\left(t,\theta,\varphi\right)=\big(r\left(t,\theta\right)\cos\varphi\;,\;r\left(t,\theta\right)\sin\varphi\,,\;z\left(t,\theta\right)\big)\,.
\label{eq:vector X after cylindrical symmetry}
\end{equation}
With this parametrization the constraint (\ref{eq:gamma 01 constraint}) reads
\begin{equation}
\dot{r}r'+\dot{z}z'=0\,,
\label{eq:gamma 01 constraint 2}
\end{equation}
where prime denotes derivative with respect to $\theta$.

Combining this result with (\ref{const2}), one obtains the following first-order system of equations \cite{Eggers:2009zz,Singularities_rel}:
\begin{align}
\dot{r} &=-z'\sqrt{\frac{1}{r'^{\,2}+z'^{\,2}}-\frac{r^{2}}{\rho^{2}}}\,, \label{eq:rdot} \\
\dot{z} &=r'\sqrt{\frac{1}{r'^{\,2}+z'^{\,2}}-\frac{r^{2}}{\rho^{2}}}\,.
\label{eq:zdot}
\end{align}

On the other hand, using the fact that $\rho$ is constant in time, the spatial components of \eqref{eq:eom} lead to
\begin{align}
\ddot{z}&=\frac{r}{\rho^{3}}\left[\left(rz''+2r'z'\right)\rho-rz'\rho'\right]\,. 
\label{eq:eom for z}\\
\ddot{r}&=\frac{r}{\rho^{3}}\left[\left(rr''+r'^{\,2}-z'^{\,2}\right)\rho-rr'\rho'\right]\,,
\label{eq:eom for r} 
\end{align}
which can also be written as
\begin{align}
\ddot z &= \frac{1}{\rho} \left(\frac{r^2 z'}{\rho}\right)' , \\
\ddot r &= \frac{1}{\rho} \left(\frac{r^2 r'}{\rho}\right)' - \frac{r}{\rho^2}\left(r'^2+z'^2\right) ,
\end{align}
where the last term in the equation for $r$ arises as an additional contribution specific to the 
$3+1$-dimensional spacetime, and is absent in the corresponding $2+1$-dimensional case. Furthermore, these two second-order equations are consistent with (\ref{eq:rdot}), (\ref{eq:zdot}), and more convenient to use for the numerical integration with a static initial condition.

\begin{figure}[t]
\centering
\includegraphics[width=\linewidth]{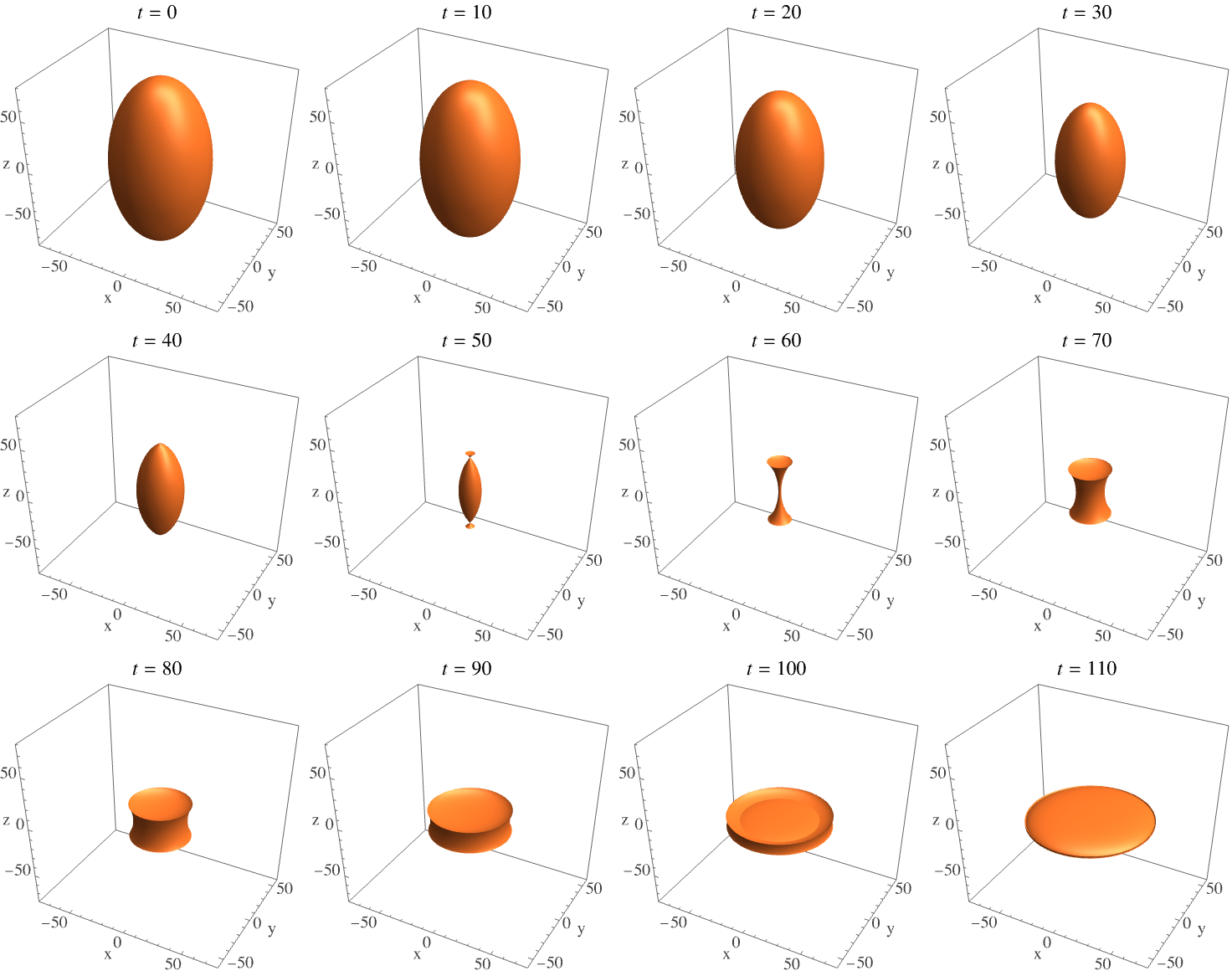}
\caption{Nambu-Goto evolution of a DW initially at rest and with a prolate ellipsoid axially symmetric shape. The polar radius is $c=80$, while the equatorial radius is $a=40$.
As the DW collapses, singular points appear on the poles, at $t\approx40$. These are axially symmetric versions of the swallowtail bifurcation for strings in $2+1$ dimensions \cite{Blanco-Pillado:2025gzs}. Due to the axial symmetry, the swallowtail singularity results in two circular cuspidal edges (one on each pole) that expand away from the symmetry axis. At a finite time, $t\approx60$, the vertical sides of the wall collide against each other in what can be seen as a almost planar domain wall collision or the collapse of a close to cylindrical wall. This results in a hyperboloid-like shape that expands in the horizontal directions, while the upper and lower caps continue traveling toward one another, producing a second wall collision at the center, $t\approx100$. The evolution continues as an expanding oblate ellipsoid with the two ring cuspidal edges at the front, which at some time, $t\approx110$, meet each other on the equator into a smooth oblate ellipsoid in an inverse swallowtail bifurcation.\\
} 
\label{fig:NG evolution}
\end{figure}

As a first instance of a non-spherical collapse, we consider the collapse of an ellipsoidal DW initially at rest, $r\left(0,\theta\right)=a\sin\theta$, $\dot{r}\left(0,\theta\right)=0$, $z\left(0,\theta\right)=c\cos\theta$, $\dot{z}\left(0,\theta\right)=0$,
which implies $\rho^2=a^{2}\sin^{2}\theta\left(a^{2}\cos^{2}\theta+c^{2}\sin^{2}\theta\right)$. 
%
%
%
We numerically solved the equations (\ref{eq:eom for z}) and (\ref{eq:eom for r}) for an initially static prolate ellipsoid with semi-minor axis $a=40$ and semi-major axis $c=80$. Here the length units are arbitrary but will be related to the DW width introduced in Sec.~\ref{Sec:FT}. Fig.~\ref{fig:NG evolution} shows several snapshots obtained from the numerical integration of this initial condition. Consistently with the results of Ref.~\cite{Singularities_rel}, the worldvolume is seen to develop singular structures within a finite time. In the following section we will analyze them in detail.

\subsection{ Cuspidal Edges in Singularity Theory }

The time evolution of the solution to the NG equation shown in Fig.~\ref{fig:NG evolution} clearly shows that there are special events where the DW surface develops singularities. These points can be understood with the tools of singularity theory as swallowtail bifurcations \cite{Eggers:2009zz,Singularities_rel}. The equivalent notion for DWs in $2+1$ dimensions was presented in \cite{Blanco-Pillado:2025gzs}. 


Although the resulting dynamical equations can be solved numerically, as we just described in the previous section \ref{Sec:Ellipsoid}, we are now interested in an asymptotic approximation of the solution near the swallowtail point $(t_*,\theta_*)$ corresponding to the first appearance of singularities. 
 
As shown in Fig.~\ref{fig:NG evolution}, there are 2 swallowtail events:  first on the pole(s) and on the equator. 
Let us start with the one on the poles.
Following~\cite{Eggers:2009zz,Blanco-Pillado:2025gzs}, we will consider an expansion of the solution near the swallowtail event that can be captured by the ansatz
\begin{align}
    r(\bar t,\bar\theta)&=\lambda_1 \bar t\bar\theta+\lambda_3 \bar\theta^3+\order{\bz^5}\label{ransatz},\\
    z(\bar t,\bar \theta) &=z_*+\lambda_0\bt+\lambda_2\bt\bz^2+\lambda_4\bz^4+\order{\bz^5},\label{ec:zansatz}
\end{align}
where the dimensions of the constants are  $[\lambda_{0,1,2}]=1$,  $[\lambda_{3,4}]=L$ 
where $\bar{t}=t-t_*$ and  $ \bar{\theta}=\theta-\theta_*$.

Inserting \cref{ransatz,ec:zansatz} into the constraint \eqref{eq:gamma 01 constraint 2}, and assuming the lightlike behaviour of the singularity, we get the following coefficient relations:
\begin{equation}
    \lambda_0=-1,\quad \lambda_2=\frac{1}{2}\lambda_1^2,
    \quad \lambda_4=\frac{3}{4} \lambda_3\, \lambda_1~,
    \label{ec:coeff-constraints}
\end{equation}
while $\lambda_{1,3}$ are left free. 
At the order displayed, the second order equations of motion do not further constrain these coefficients, but instead determine the higher order terms omitted from the local ansatz. This agreement can, in principle, be systematically extended to higher orders by incorporating additional terms in the expansion introduced above that capture the timelike nature of the worldvolume around the singular point.

Further, with the rescalings  $\vartheta=\lambda_1\,\bz$ and $\tau=\bt/\eta$ with $\eta=\lambda_3/(4\,\lambda_1^3)$, we end up with
\begin{align}
 \frac{r}{\eta}&=\tau\vartheta+4\vartheta^3\label{r_swallow}\\
    \frac{z-z_*}{\eta}&=-\tau+\frac{1}{2}\tau\vartheta^2+3\vartheta^4\label{z_swallow}
\end{align}
The resulting asymptotic expansion therefore takes the universal swallowtail form~\cite{Arnold1975,Eggers:2009zz}, consistently with the structure previously identified in $2+1$ dimensions~\cite{Eggers:2009zz,Blanco-Pillado:2025gzs}. In the present case, the only difference is the presence of rotational symmetry around the axis.\footnote{This is a consequence of the fact that the additional contributions specific to the $3+1$-dimensional problem vanish at leading order in the ultrarelativistic limit~\cite{Eggers:2009zz}.}

\begin{figure}[t]
\centering
\includegraphics[width=\linewidth]{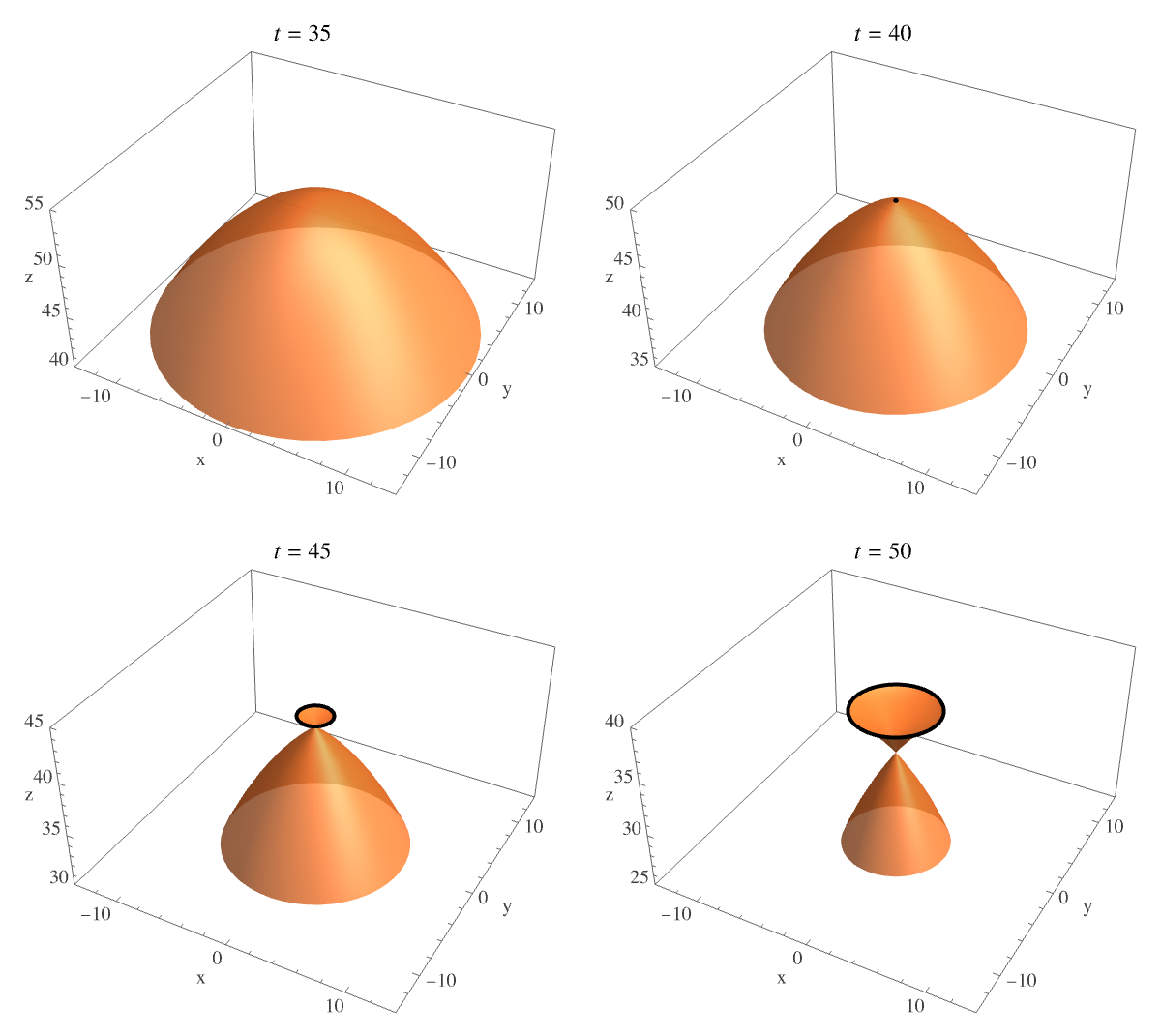}
\caption{Zoomed-in view of the region near the pole of the axially symmetric collapsing domain wall shown in Fig.~\ref{fig:NG evolution}. The black line indicates the formation and subsequent evolution of the cuspidal edge singularity, which propagates rapidly downward along the $z$ direction. Note that he range of the $z$ axis has been adjusted between panels to improve the visualization of the relevant structure.
}
\label{fig:Trumpet}
\end{figure}

Similarly to the $2+1$-dimensional case, the bifurcation gives rise to a novel structure at the poles. In particular, the evolution leads to the formation of a trumpet like surface in the vicinity of the self-intersection point (see Fig. \ref{fig:Trumpet}). This surface is characterized by the presence of a circular cuspidal edge singularity, which expands and propagates toward decreasing values of the $z$ coordinate. These cuspidal edges are line-like singular structures along which the domain wall attains the speed of light. In directions transverse to the edge, the local geometry exhibits the characteristic profile of a regular cusp, which motivates the terminology “cuspidal edge”.


The other type of axially symmetric swallowtail singularity is extended along the equator, when the wall is oblate. The asymptotic expansion near the singularity $z=0$ is again of the form \eqref{ec:zansatz} with the role of the coordinates $r$ and $z$ reversed with respect to \cref{r_swallow,z_swallow}, hence the locus  of points that develop a cuspidal edge coincide with the equatorial circle. 

Notice that in this case both the merging cuspidal edge and the swallowtail event are circles, so the induced metric on the wall has only one vanishing eigenvalue. Instead, the swallowtail events on the poles are pointlike and there $g_{ab}$ is doubly degenerate. 

\section{Eikonal approximation}
\label{sec:EikoalApp}

\begin{figure}[t]
    \centering
        \includegraphics[width=0.9\linewidth]{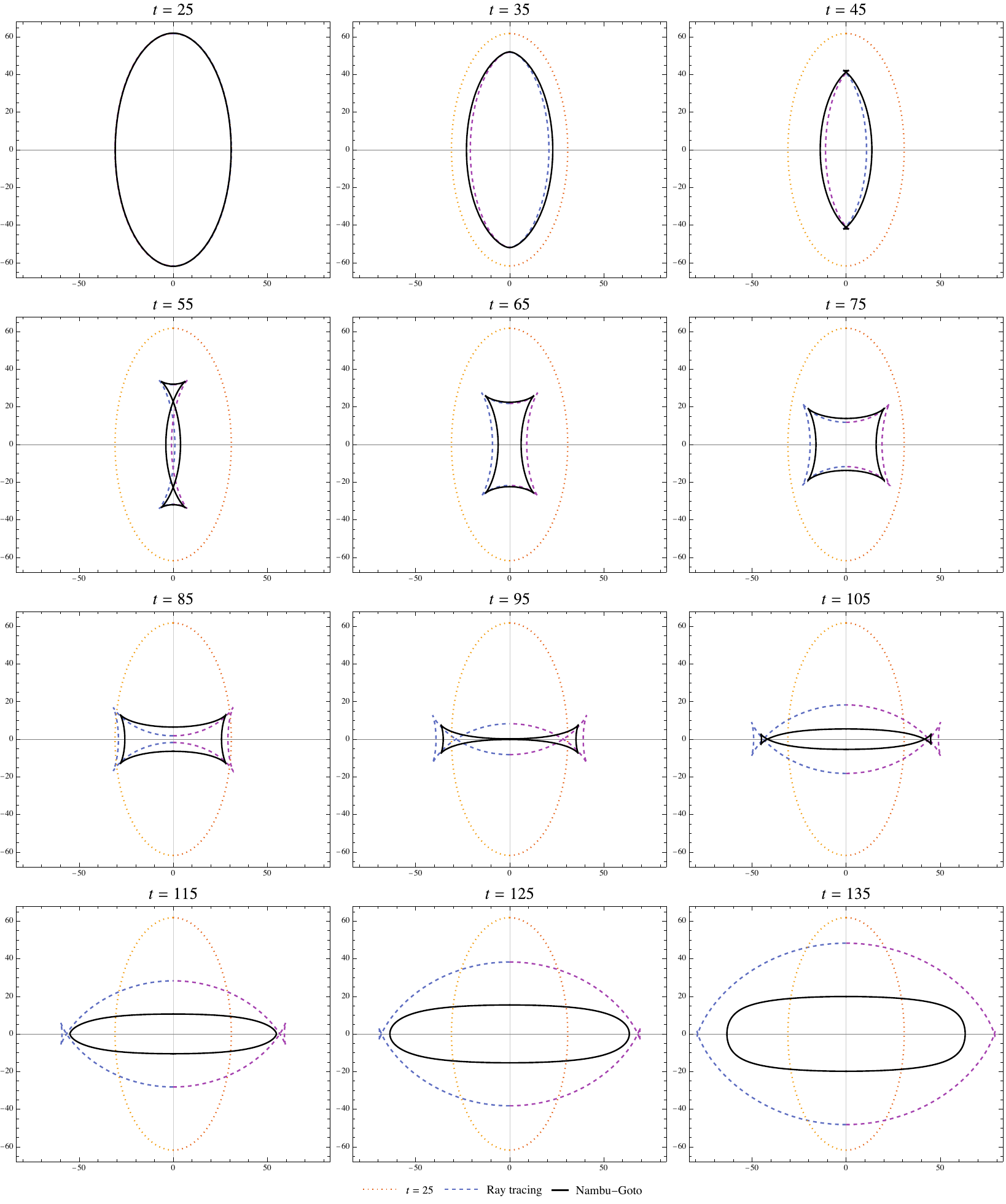}
    \caption{Comparison of the Nambu-Goto evolution and ray tracing evolution for the prolate spheroid in Fig. \ref{fig:NG evolution}. Here we present the transversal section of the spheroid on the plane $y=0$. The black continuous line corresponds to the Nambu-Goto solution parametrized by Eq. (\ref{eq:vector X after cylindrical symmetry}). The blue dashed line, on the other hand, is the eikonal approximation given by  Eq. (\ref{eq:eikonal approximation eq}). As a reference, the dotted orange line corresponds to the shape of the DW at $t=25$.
}
    \label{fig:comparison}
\end{figure}

Let us now examine how, for collapsing domain walls, the Nambu–Goto equations \eqref{flow} admit an approximate description that leads to an enormous simplification of the dynamics \cite{Hoppe:1995sn,Eggers:2009zz,Singularities_rel}.
The main observation is that in the cases of physical interest (closed DWs with large overall size compared to its thickness) the DW necessarily reaches (ultra-)  relativistic velocities along the evolution -- even before the DW shape changes much. More precisely, the local velocity of the wall elements, $|\dot{\mathbf{x}}|$, approaches unity over a sizeable portion of the wall.
This simple and quite robust fact (for closed DWs) allows to find an approximate global solution to the DW motion. 

Setting $|\dot{\mathbf{x}}_0|=1$  (hence, $V=1$) as a zeroth order approximation into the NG equation \eqref{flow} leads to
\begin{equation}
    \dot{\mathbf{x}}=\mathbf{n}~,
\end{equation}
where, again, $\mathbf{n}$ is the inward normal vector at any point, which in principle could still depend on time. This is the eikonal equation, used to trace the light rays in geometrical optics (with constant refraction index). In a sense, once the wall enters  the ultrarelativistic limit, it behaves as a light `wavefront'.

The evolution simplifies even further in this limit, since at leading order (in $1-|\mathbf{\dot x_0}|$) the normal vector $\mathbf{n}$ gets frozen. Indeed it is easy to see that in full generality the evolution equation for $\mathbf{n}$ is
\begin{equation}\label{nflow}
    \dot{\mathbf{n}} = \partial_a V \, g^{ab}\, \partial_b \mathbf{x}~. 
\end{equation}
This follows  by differentiating $\mathbf{n}\cdot \partial_b \mathbf{x}=0$ and using \eqref{flow} (evolution equations for various geometrical quantities like the spatial metric $g_{ab}$ can be obtained similarly \cite{Hoppe:1995sn}). In the (everywhere) ultrarelativistic limit, $|\dot{\mathbf{x}}_0|=1$, the magnitude of the speed remains $V=1$ (see \eqref{V}) so indeed $\dot{\mathbf{n}} =0$.

Then, a very simple approximate solution emerges of the form of a simple ray-tracing
\begin{equation}
\label{eq:eikonal approximation eq}
    \mathbf{x}(t,\sigma^a)=\mathbf{x}(0,\sigma^a)+\mathbf{n}(0,\sigma^a)t \qquad {\rm (ultrarelativistic~approximation)}.
\end{equation}
This gives a global parameterization of the whole DW in terms only of the shape of the initial condition -- which must be taken when the DW is sufficiently relativistic. Let us note also that 
\eqref{eq:eikonal approximation eq} actually solves exactly the constraint \eqref{eq:gamma 01 constraint}. 



The ray tracing approximation \eqref{eq:eikonal approximation eq} holds so much potential that it is worth testing its limits. The global evolution from \eqref{eq:eikonal approximation eq} matches the actual NG evolution remarkably well for the collapsing domain walls we are considering here. 
The good qualitative agreement can be seen in Fig.~\ref{fig:comparison} 
, which contains snapshots of the evolution with the two methods.

To compare the two evolutions, we first determine numerically the time at which the Nambu–Goto evolution enters the relativistic regime. This is achieved by extracting the velocity of the poles from the numerical solution; we find that by $t=25$, the velocity has already reached $v=0.99$. At this stage, the shape of the domain wall is well approximated by an ellipsoid with polar radius $a=61.8$ and equatorial radius $c=30.9$. This configuration is then adopted as the initial condition for the ray-tracing (eikonal) approximation. The subsequent evolution obtained from the exact Nambu–Goto dynamics and from the eikonal approximation is compared from this time onward in Fig.~\ref{fig:comparison}.

The ultrarelativistic eikonal limit reproduces precisely the singularity structure of the domain wall collapse, as it must since these points move with $|\dot{\mathbf{x}}|=1$ in NG. Furthermore, this approximation is expected to continue being accurate throughout the collapse phase, that is, until the wall bounces back to recollapse after a time 
order $\Delta t \sim 2 |\mathbf{x_0}|$, because the real wall slows down before recollapsing again. In the field theory simulations, though, the DW emits radiation to infinity before recollapsing (if at all). It is intriguing whether \eqref{eq:eikonal approximation eq} can capture the shape of this real radiation front to infinity. We leave this question for future work.


In the following subsection, we demonstrate that the eikonal approximation accurately reproduces the same singularity structure, including the formation of cuspidal edges, for the same class of simple initial configurations that we discussed above. This level of agreement is not unexpected, as the approximation should perform optimally in precisely those regions where the Nambu–Goto dynamics becomes ultrarelativistic, with velocities approaching $V=1$.

\subsection{Singularity formation from ray tracing}
\label{sec:formation-eikonal}

We first focus on the prolate ellipsoidal domain wall, generated by revolution of the ellipse
\begin{equation}
    r(\theta)=\frac{a}{\sqrt{1-\epsilon^2 \cos^2{\theta}}}
\end{equation}
around the semi-major axis
where $a$ is the length of the semi-minor axis and $0<\epsilon<1$ is the eccentricity.

The worldvolume of the wall in the eikonal approximation is  parametrized by 
\begin{equation}
    X^\mu=\Big(t\;,\; \mathbf{x}(0,\theta,\varphi)+\mathbf{n}(0,\theta,\varphi)\;t \Big), 
    \label{eq:Xmu}
\end{equation} 
with 
\begin{align}
    \mathbf{x}(0,\theta,\varphi)&=\qty[\frac{a \sin{\theta} \cos{\varphi}}{\sqrt{1-\epsilon^2  \cos^2{\theta}}},\frac{a \sin{\theta} \sin{\varphi}}{\sqrt{1-\epsilon^2  \cos^2{\theta}}},\frac{a\cos{\theta}}{\sqrt{1-\epsilon^2  \cos^2{\theta}}}], \\[2mm]
    \mathbf{n}(0,\theta,\varphi)&=-\frac{1}{ \sqrt{(\epsilon^2 -2) \epsilon^2  \cos^2{\theta}+1}}\qty[\sin{\theta} \cos{\varphi},\sin{\theta} \sin{\varphi},(1-\epsilon^2 ) \cos{\theta}]\,,\notag
 \end{align}
defining the initial position of the wall and its unit normal vector (pointing inwards) at $t=0$.

As argued, the surface given by the spatial part of \eqref{eq:Xmu} develops a singularity at a finite value of the time coordinate $t=t_*$. By symmetry, such a point must occur at the revolution axis, $i.e.$ $\theta_*=0$. Without loss of generality, we may also restrict to the $\varphi=0$ plane. In that case, we may expand the solution \eqref{eq:Xmu} around $\theta=0$ to find
\begin{align}
    r&= \left(\frac{a}{\sqrt{1-\epsilon^2 }}+\frac{t}{\epsilon^2 -1}\right)\theta +\frac{ a(2 \epsilon^2 +1) (1-\epsilon^2 )^{3/2}+t (2 (\epsilon^2 -2) \epsilon^2 -1)}{6 (\epsilon^2 -1)^3}\theta ^3+\order{\theta ^5}\label{eq:r_eikonal}\\[2mm]
    z&=\frac{a}{\sqrt{1-\epsilon^2 }}-t+\frac{\left(a (\epsilon^2 -1)+t \sqrt{1-\epsilon^2 }\right)}{2 (1-\epsilon^2 )^{5/2}}\theta ^2 +\frac{a (8 \epsilon^2 +1) (1-\epsilon^2 )^{3/2}+t (8 (\epsilon^2 -2) \epsilon^2 -1)}{24 (\epsilon^2 -1)^4}\theta ^4 +\order{\theta ^5}\label{eq:z_eikonal}
\end{align}

From the above expansion, one obtains the time $t_* =a \sqrt{1-\epsilon^2 }$ at which the singularity develops, from the condition that the linear term in \cref{eq:r_eikonal} vanishes.
Keeping only the lowest nontrivial orders in $\bar{t} = t-t_*$, the expansion near the singular time reads
\begin{align}
    r(t,\theta)
    &=
    -\frac{\bar{t}}{1-\epsilon^2}\,\theta
    +
    \frac{\epsilon^2 t_*}{2(1-\epsilon^2)^3}\,\theta^3
    +\cdots ,
    \\[2mm]
    z(t,\theta)
    &=
    z_*
    -
    \bar{t}
    +
    \frac{\bar{t}}{2(1-\epsilon^2)^2}\,\theta^2
    -
    \frac{3\epsilon^2 t_*}{8(1-\epsilon^2)^4}\,\theta^4
    +\cdots .
\end{align}
which after an irrelevant shift in the $z$ direction, coincides with \cref{ransatz,ec:zansatz} together with the coefficient relations given by \cref{ec:coeff-constraints}, \emph{i.e.} the evolution using the eikonal approximation of the initial conditions leads to the formation of the swallowtail singularity. 
Furthermore, this local expansion of the worldvolume also predicts the formation of the cuspidal edges we saw earlier. This result is not surprising. The key point is that the eikonal approximation captures the essential phenomenology of collapsing domain walls: regions with different curvature evolve unevenly, leading to the formation of caustics and hence to singularities in the domain wall worldvolume. Our analysis shows that the eikonal evolution of the initial configuration gives rise to this class of singularities and, moreover, allows the coefficients of the local worldvolume expansion around the singular points to be determined directly from the parameters of the initial data. As discussed above, the full numerical Nambu--Goto evolution exhibits qualitatively the same type of singular behaviour. This is in agreement with the general expectations of singularity theory, according to which the local type of a generic singularity is structurally stable. Therefore, even if the approximate dynamics does not necessarily reproduce the exact Nambu–Goto motion in full detail and leads to slightly different values of the parameters characterizing the singular configuration, the nature of the singularity itself is expected to be preserved. 

Encouraged by the faithfulness of this ray racing (or eikonal) approximation, we use it to assess the genericity of the formation of cuspidal singularities in the next section.

\section{Genericity}
\label{genericity}

We will now argue that the formation of cuspidal singularities on domain walls is a generic feature of their collapse. More precisely, for smooth initial configurations that are otherwise generic, i.e. not specially fine-tuned, cuspidal singularities are expected to form dynamically. Our argument relies on the eikonal approximation to the domain wall dynamics, within which one finds that an arbitrary closed domain wall, regardless of its initial shape, generically undergoes several catastrophic singular events at finite time during its collapsing phase\footnote{The situation is reminiscent of the classical result that cusps arise generically in smooth closed cosmic string loops~\cite{Vilenkin:2000jqa}.}.\\

It is important to  distinguish between two very different types of cuspidal singularities that we will encounter generically in the domain wall collapse:

\begin{enumerate}
    \item \textbf{\textit{Cuspidal Edge} Singularities}, exemplified by those visible in Fig.~\ref{fig:NG evolution} and Fig.~\ref{fig:Trumpet}
, which are extended over a finite length at each instant of time and propagate along the worldvolume for a finite interval, moving at the speed of light in the ambient spacetime. 
    
    \item \textbf{\textit{Cuspidal Vertex} Singularities}, which we introduce below, for which the singular set takes the form of a pyramidal spike and, unlike the previous case, occurs only as an instantaneous singular event.
\end{enumerate}

There is a genericity argument for both of them. However, the physical impact of  cuspidal edges ({\em e.g.} in a GW signal) is expected to be larger because of their longer and larger presence.  

\subsection{Cuspidal Edges}

It is not difficult to understand why cuspidal edges, i.e. line-like singularities with a transverse cusp profile, are generic and, for smooth non-spherical initial configurations, essentially unavoidable. As the collapse develops, the wall quickly attains relativistic velocities, so that the ray-tracing approximation provides an accurate description of the evolution. Because the initial shape is not spherical, there necessarily exist regions whose local radius of curvature is smaller than the overall characteristic size of the domain wall. In the ray-tracing equation \eqref{eq:eikonal approximation eq}, this local radius of curvature determines the locus at which neighboring rays focus. Such focusing generically gives rise to line-like singularities, namely the cuspidal edges. The remaining question is therefore to determine the generic local shape with which these domain wall cuspidal edges are formed.

\begin{figure}[t]
    \centering
        \includegraphics[width=\linewidth]{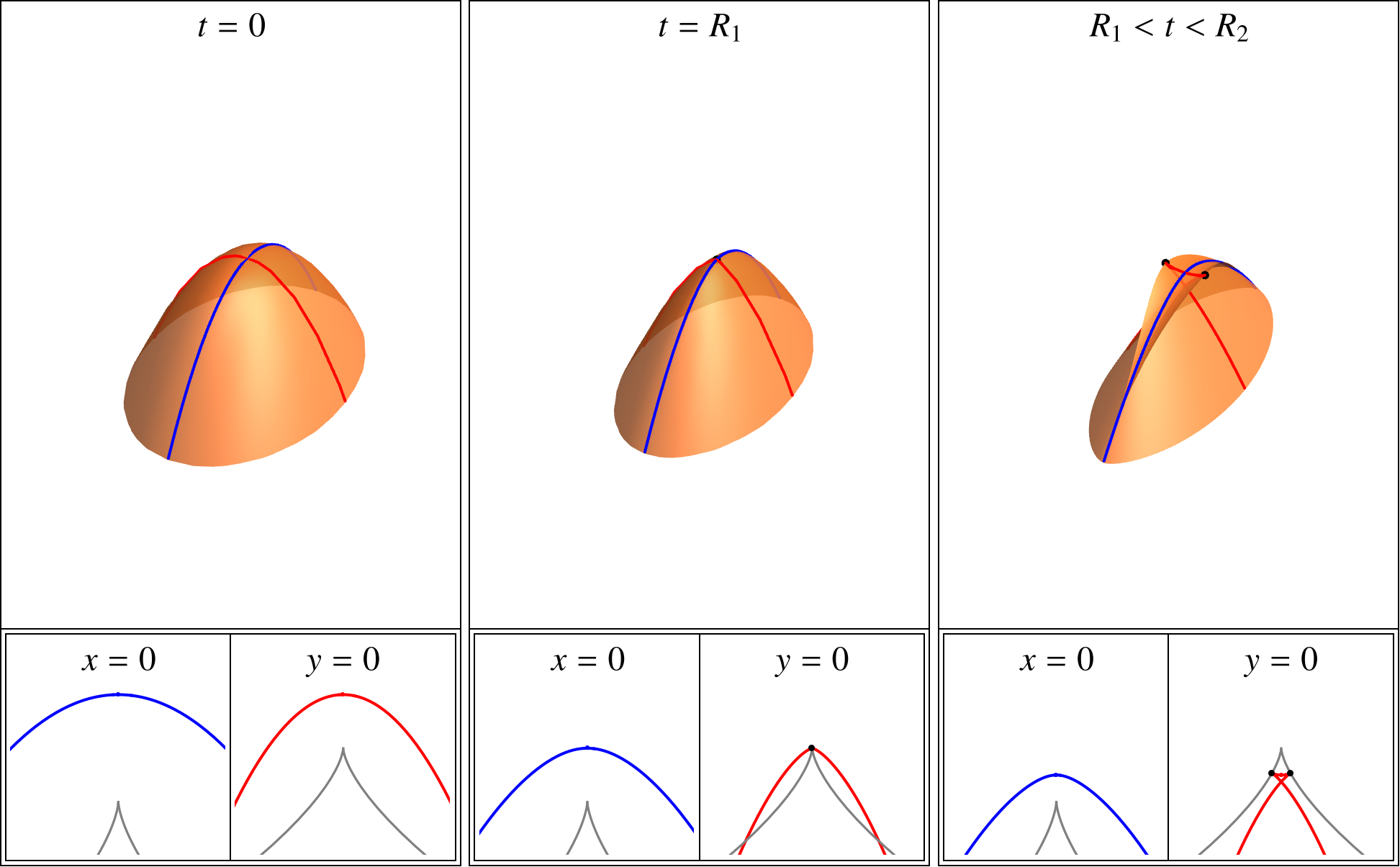}    
    \caption{Formation of cuspidal edges in the generic case (no axial symmetry assumed). The snapshots are taken from the eikonal time evolution of a paraboloid with principal curvatures $\kappa_1$ and $\kappa_2$. The solid blue and red lines show the evolution of the points along the principal axes  $x=0$ and $y=0$, respectively. Observe that at $t=R_1$, the evolution develops a swallowtail singularity along the $y=0$ axis. The singularity unfolds as shown in the right panel: two cuspidal edges form and separate, leaving behind a characteristic `lip' structure. The black dots represent the points of intersection of the cuspidal edges with the axes $x=0$, $y=0$. In gray, we present their trajectories.}
    \label{fig:eikonal_paraboloidA}
\end{figure}

The  generic case can be illustrated as follows. Any smooth surface can be locally approximated by a tangent (sometimes called `osculating') paraboloid. For any given point $p$ of an arbitrary smooth surface $\Sigma$, and for a sufficiently small neighborhood of $p$ parametrized by the $x$ and $y$ coordinates, the surface can be defined by the set of points that satisfy  $z=f(x,y)$. Moreover, choosing $p$ as the origin of a local coordinate system, the $z$ axis in the direction of the normal to the surface at $p$, and appropriate basis vectors, the surface can be locally approximated by 
\begin{equation}
    z=\frac{1}{2}\Big(\kappa_1\,x^2+\kappa_2 \, y^2\Big)+\order{x^3,y^3} 
    \label{oscul-par}
\end{equation}
with principal curvatures 
\begin{equation}
    \kappa_1=-\frac{1}{R_1} \qquad \kappa_2=-\frac{1}{R_2}~,
\end{equation}
and $R_{1,2}$ are the principal curvature radii at the point $p$. We shall take $R_1<R_2$.

It is illustrative, then, to  consider the eikonal evolution of a generic paraboloid as the surface representing locally a section of a domain wall collapsing. 
The time evolution is given by  \eqref{eq:eikonal approximation eq}, with the initial condition
\begin{equation}
    \mathbf{x}(0,\sigma_1,\sigma_2)=\left\{\sigma_1,\sigma_2,\frac{1}{2}\Big(\kappa_1 \sigma_1^2+\kappa_2 \sigma_2^2\Big)\right\}~.
    \label{oscul-par2}
\end{equation}
The curvature signature is chosen consistently with a convex orientation, meaning $\mathbf{n}(0,0,0) = -\hat{z}$.

In the eikonal (ultrarelativistic) approximation \eqref{eq:eikonal approximation eq}, the full parameterized worldvolume is
\begin{align}
    x &= \sigma _1 \left(1-\frac{\kappa_1 t}{\sqrt{\kappa _1^2 \sigma_1^2+\kappa _2^2 \sigma
   _2^2+1}}\right)~,\\
   y &= \sigma _2 \left(1-\frac{\kappa_2 t}{\sqrt{\kappa _1^2 \sigma_1^2+\kappa_2^2 \sigma_2^2+1}}\right)~,\\
   z &= -\frac{1}{2} \kappa_1 \sigma _1^2-\frac{1}{2} \kappa_2 \sigma _2^2-\frac{t}{\sqrt{\kappa_1^2 \sigma_1^2+\kappa _2^2 \sigma_2^2+1}}~.
\end{align}

\begin{figure}[t]
    \centering
        \includegraphics[width=\linewidth]{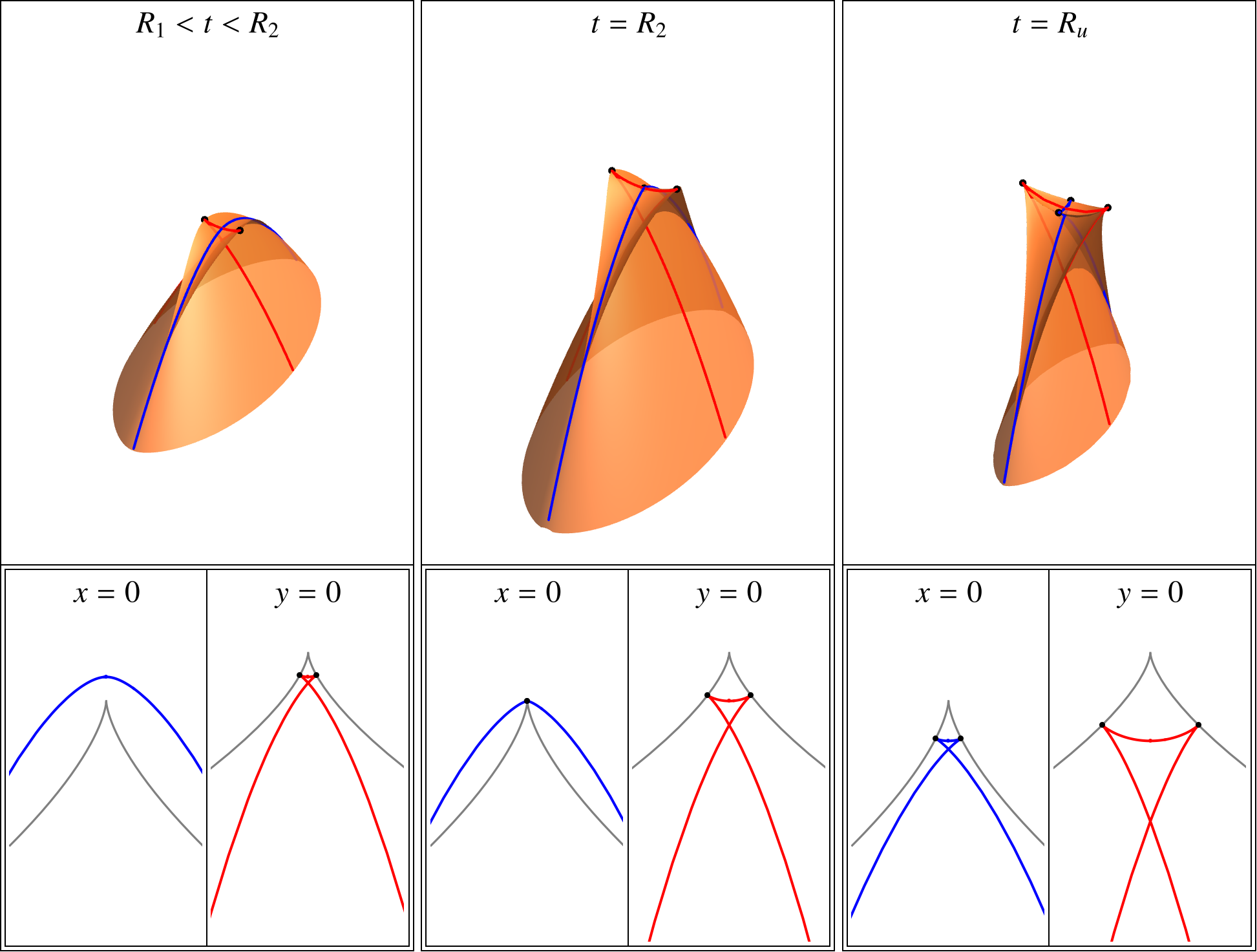}
    
    \caption{
    Eikonal time evolution of a paraboloid with principal curvatures $\kappa_1$ and $\kappa_2$. The solid blue and red lines show the evolution of the points along the principal axes $x=0$ and $y=0$, respectively. Observe at $t=R_2$ the development of the swallowtail singularity along the $x=0$ axis. The right panel shows the two cuspidal vertices that appear at $t=R_u=R_2\left({R_2}/{R_1}\right)^{1/2}$, the focusing time associated with the paraboloid's umbilical points. The black dots represent the points of intersection of the cuspidal edges with the planes $x=0$, $y=0$. In gray, we present their trajectories.}
    \label{fig:eikonal_paraboloidB}
\end{figure}

Fig.~\ref{fig:eikonal_paraboloidA} and Fig.~\ref{fig:eikonal_paraboloidB} show different time frames of the evolution under this eikonal approximation. It is clear that different singularities develop. 
Looking first at Fig.~\ref{fig:eikonal_paraboloidA}, we find a first singularity forming at $t=R_1$, that  is recognized as a swallowtail event that subsequently results in the formation of two cuspidal edges. In contrast with the axially symmetric case, shown in Fig. \ref{fig:Trumpet}, the cuspidal edge in this case is not a circle but a pair of line-like cusps that merge smoothly at a finite aperture (that grows with time). The two cuspidal edges are joined by a lip-like structure that grows both in the $x$ and the $y$ directions. In singularity theory notation, the cuspidal edges and swallowtail event correspond to singularities of type $A_2$ and $A_3$ respectively, see e.g.~\cite{Arnold1975,iiasa8095}.

The main conclusions of the discussion thus far are the following. First, cuspidal edges form irrespective of whether or not the initial configuration is axially symmetric. Second, the third snapshot shown in Fig.~\ref{fig:eikonal_paraboloidA} represents the \emph{generic} local shape with which domain wall cuspidal edges are generated during collapse. This expectation is robust, since it relies only on the fact that the focusing points on the initial surface, which are responsible for the swallowtail formation events, generically possess distinct principal curvatures $\kappa_i$.

Similar to the expansion \cref{eq:r_eikonal} and \cref{eq:z_eikonal}, it is possible to expand the eikonal evolution around the singularity at $t=R_1$. We find
\begin{align}
    x &= - \kappa _1 \sigma _1 \tau +\frac{1}{2} \kappa _1 \sigma _1 \tau  \left(\kappa _1^2 \sigma _1^2+\kappa _2^2 \sigma _2^2\right)+\frac{1}{2} \sigma _1 \left(\kappa_1^2 \sigma _1^2+\kappa _2^2 \sigma _2^2\right)+ ...,\nonumber\\
     y &= -\kappa _2 \sigma _2 \tau +\frac{1}{2} \kappa _2 \sigma _2 \tau  \left(\kappa _1^2 \sigma _1^2+\kappa _2^2 \sigma_2^2\right)
     +\frac{1}{2} \frac{\kappa_2}{\kappa_1} \sigma _2 \left(\kappa_1^2 \sigma _1^2+\kappa _2^2 \sigma _2^2\right)
     +\left(1-\frac{\kappa _2}{\kappa _1}\right)\sigma _2+ ...,\label{319}\ \\
    z &=-\tau 
    +\frac{1}{2} \tau \left(\kappa _1^2 \sigma _1^2+\kappa _2^2 \sigma _2^2\right)    
    -\frac{1}{2}\left(1-\frac{\kappa _2}{\kappa _1}\right) \kappa _2 \sigma _2^2-\frac{3}{8\kappa _1}\left(\kappa _1^2 \sigma_1^2+\kappa _2^2 \sigma _2^2\right)^2+ ..., \nonumber
\end{align}
where $\tau=t-R_1$. 
This extends to the generic, non-axially symmetric case the local expansion of the swallowtail event that leads to the formation of the cuspidal edges. One can readily see that for  $\kappa_1=\kappa_2=\kappa$, this expansion reproduces the axially symmetric swallowtail \cref{r_swallow,z_swallow}.

Since we are assuming $R_1<R_2$, one can think that first singularity forms by focusing rays labeled by $\sigma_1$, so near $\sigma_2=1$ one basically has the same swallowtail as in $2+1$ \cite{Blanco-Pillado:2025gzs} trivially extended in $\sigma_2$.
The last term in \eqref{319} captures how at some certain aperture (at some point in the  $\sigma_2$ direction) the two cuspidal edges merge smoothly.

\subsection{Cuspidal Vertex }

The evolution of the initial paraboloid shape also reveals another surprise. The initial shape contains another parameter, the smaller curvature scale $|\kappa_2|<|\kappa_1|$. Clearly, there has to be another focusing event associated to it, which happens later on.  
The eikonal evolution guarantees that another swallowtail event has to occur at $t=R_2$. Indeed, as seen in the central panel of Fig.~\ref{fig:eikonal_paraboloidB}, a second pair of lips forms, with the same development except that it occurs on top of the previously formed lip structure (which already contains its own cuspidal edge). For $t>R_2$, the DW surface contains four cuspidal edges.

Shortly after $t=R_2$, a pair of special events appear in which the four edges converge to a point to form a spike structure -- this is what we call a cuspidal vertex. As in cosmic string cusps in $3+1$ dimensions, this is an instantaneous event. In singularity theory language, this is a higher order catastrophe, known as the $D_4^+$, or \emph{hyperbolic umbilic} catastrophe~\cite{Arnold1975,poston1996catastrophe}.

Indeed, this singularity occurs at $t = R_1\left({R_2}/{R_1}\right)^{3/2} = R_2\left({R_2}/{R_1}\right)^{1/2}> R_2$, 
which is precisely the curvature radius associated with the umbilical points located near the center of the paraboloid. Fig.~\ref{fig:eikonal_paraboloidB} shows the time evolution from the final frame of Fig.~\ref{fig:eikonal_paraboloidA} to the formation of this new type of singularity at the time predicted earlier. At the tip of the resulting spike, four cuspidal edges
meet at a single point; the cuspidal vertex. This strongly suggests that the subsequent evolution of the umbilical points on the initial surface gives rise to the cuspidal vertex singularity visible in the final frame of Fig.~\ref{fig:eikonal_paraboloidB}. In what follows, we establish this explicitly\footnote{Although the profiles shown here have been obtained within the ultrarelativistic eikonal approximation, the same local singularity structure is expected to arise in the evolution governed by the full Nambu--Goto equations.}.\\

We begin by constructing the most general local parametrization of an umbilic point located at the origin of our coordinate system. Because the two principal curvature radii coincide at an umbilic point, the quadratic approximation \eqref{oscul-par} does not suffice to characterize its local geometry. It is therefore necessary to extend the local expansion to a higher order. The first nontrivial contribution is therefore furnished by the most general cubic form appearing in the Taylor expansion of the surface about the umbilic point. Accordingly, the local form of the surface in the vicinity of the umbilic can be written as

\begin{equation}
    z=\frac{1}{2}\kappa(x^2+y^2)+\frac{1}{6}\qty(\alpha x^3+3\beta x^2y+3\gamma xy^2+\delta y^3)+\order{x^4,y^4}, \label{umb}
\end{equation}
where $\alpha, \beta, \gamma, \delta$ are coefficients that determine the type of umbilical point.

In the remainder of this work, we shall consider triaxial ellipsoids, as good enough representatives of  generic non-spherical shapes. It can be shown that for generic triaxial ellipsoids the umbilical points are of hyperbolic type, and upon a redefinition of axes, the local surface can be further reduced to \cite{umbilics3space}
\begin{equation}
    z=\frac{1}{2}\kappa (x^2+y^2)+\frac{1}{6}Cx(x^2+y^2)+\order{x^4,y^4},
    \label{local_umbilic}
\end{equation}
with $C$ a constant.

\begin{figure}[t!]
\centering
\includegraphics[scale=0.34]{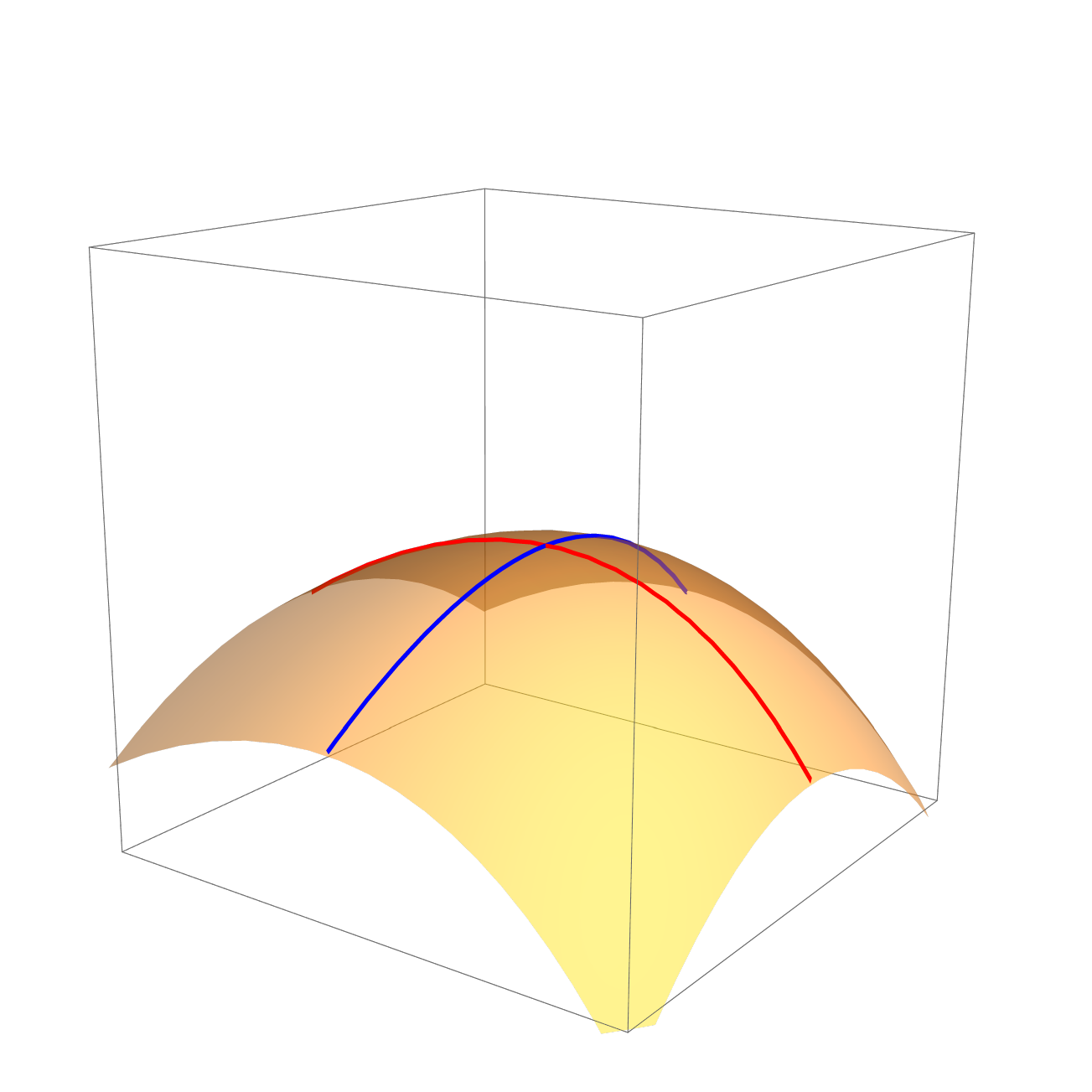}
\includegraphics[scale=0.34]{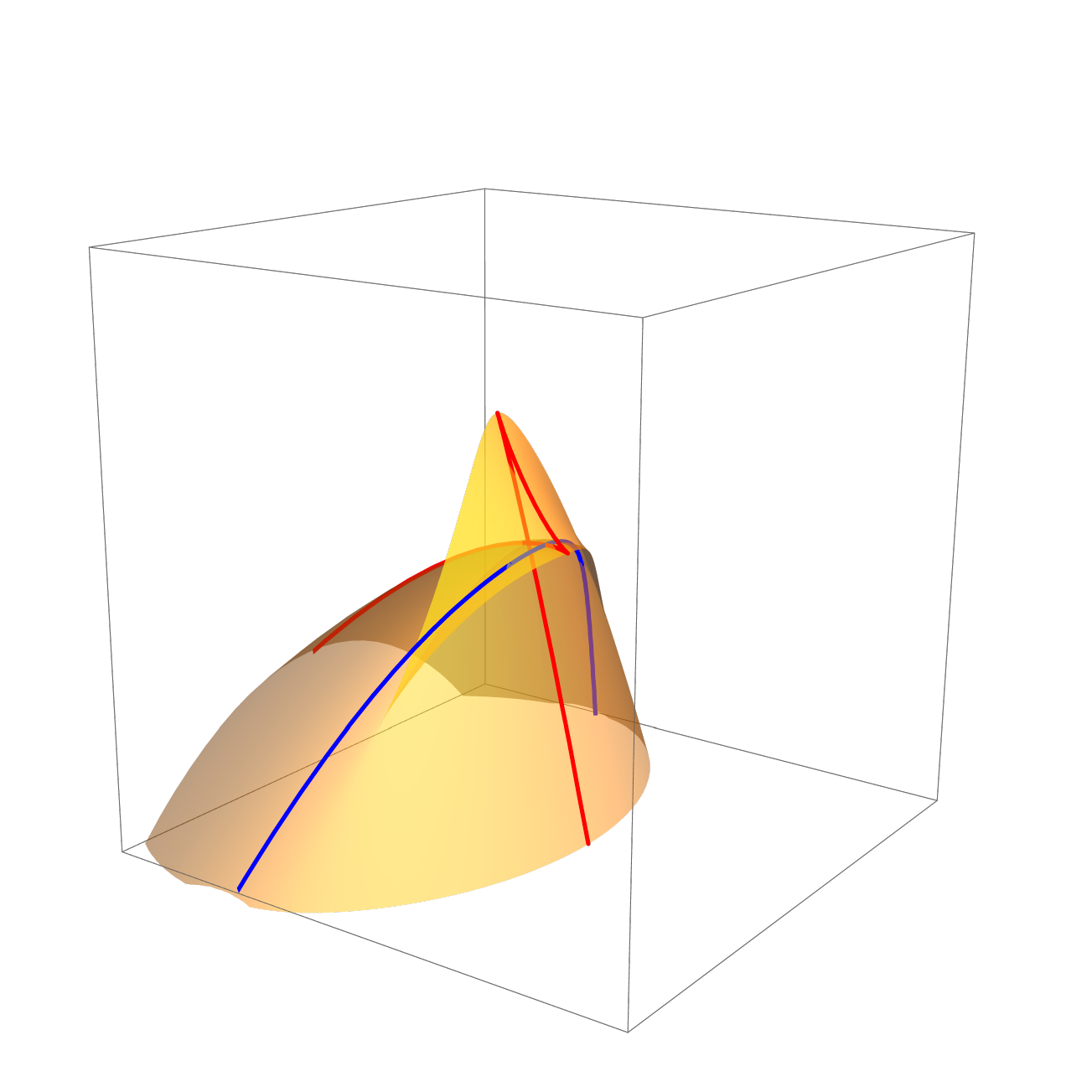}
\includegraphics[scale=0.5]{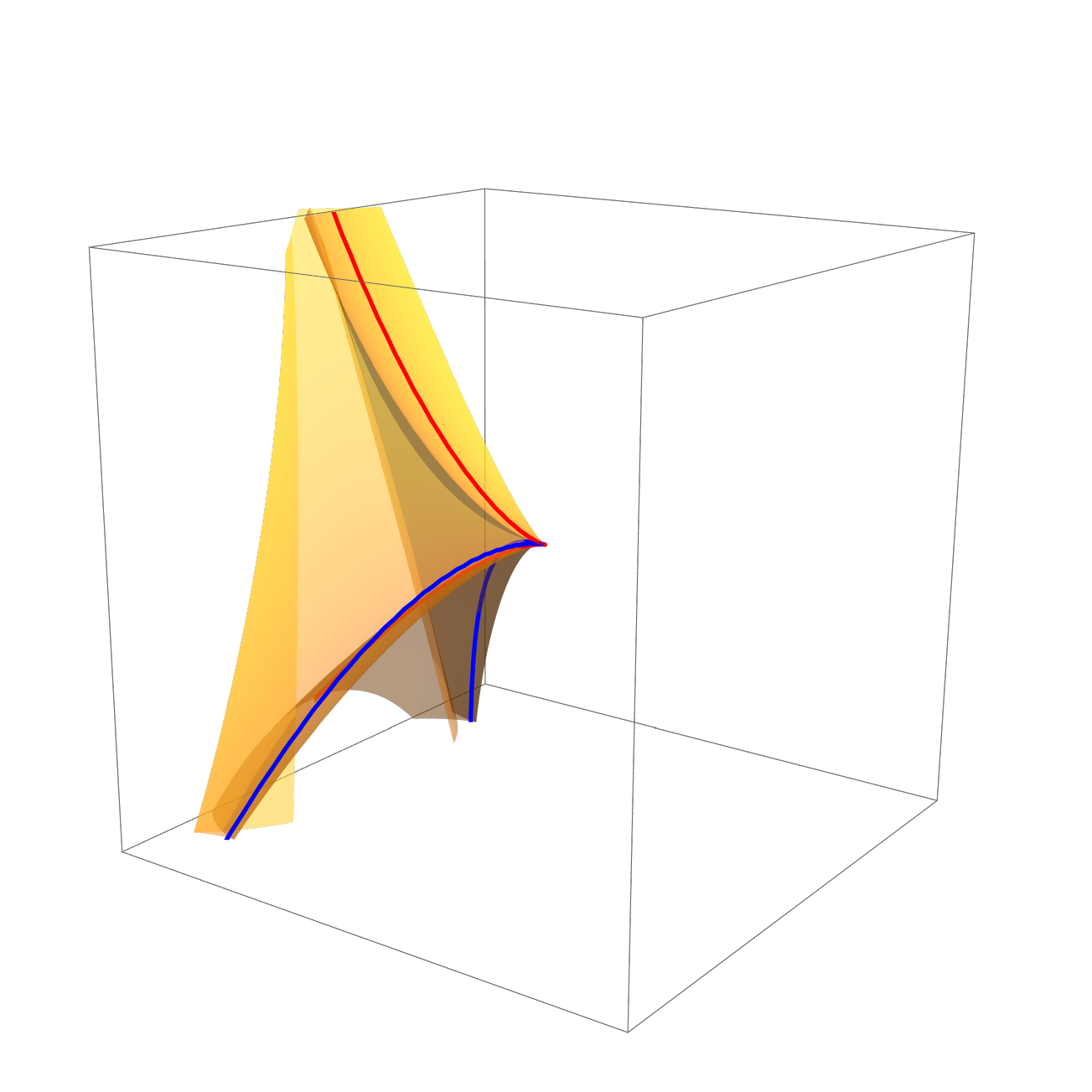}
\caption{ Eikonal development of the surface \eqref{local_umbilic}, starting at the top left. A local parametrization of these surfaces is given by \cref{X_umbilic,Y_umbilic,Z_umbilic}.}
\label{fig:NG evolution-Umbilic}
\end{figure}
Now, one can obtain an approximate description of the shape of the wall's worldvolume in a neighborhood of the singularity event via the eikonal development of \cref{local_umbilic}. Indeed, let $\mathbf{x}(\sigma^i)=(\sigma_1,\sigma_2,z(\sigma_1,\sigma_2))$ be the local parametrisation of the wall around an umbilical point. Then, the associated worldvolume in the eikonal approximation is given by \eqref{eq:Xmu}.
Now, defining $\tau=t-t_*$, with $t_*=\kappa^{-1}$ the time of catastrophe formation, we may write the spatial components of the wall worldvolume up to $\order{(\sigma^i)^4}$:
\begin{align}
    x&=\kappa  \tau  \sigma_1-\frac{C (1-\kappa  \tau )}{6 \kappa } \left(3 \sigma_1^2+\sigma_2^2\right)+\frac{1}{2} \kappa ^2 (1-\kappa  \tau ) \sigma_1 \left(\sigma_1^2+\sigma_2^2\right)+...~, \label{X_umbilic}\\[2mm]
    y&=\kappa  \tau  \sigma_2-\frac{C (1-\kappa  \tau ) \sigma_1 \sigma_2}{3 \kappa }-\frac{1}{2} \kappa ^2 (\kappa  \tau -1) \sigma_2 \left(\sigma_1^2+\sigma_2^2\right)+...~,\label{Y_umbilic}\\[2mm]
    z&=\frac{1}{\kappa }-\tau+\frac{1}{6} \left(C (3 \kappa  \tau -2) \sigma_1+3 \kappa ^2 \tau \right)\left(\sigma_1^2+\sigma_2^2\right)+... ~,\label{Z_umbilic}
\end{align}
see Fig.~\ref{fig:NG evolution-Umbilic}. This evolution process was already described by Arnold (see Fig. 5 of \cite{Arnold1975}). Up to a constant shift of the $z$ coordinate and a simultaneous reversal of the orientations of the $\tau$ and $z$ coordinates, this expansion agrees, for $C=0$, with \eqref{319} (with $\kappa_1=\kappa_2$).

\begin{figure}[t]
    \centering
        \includegraphics[width=0.7\linewidth]{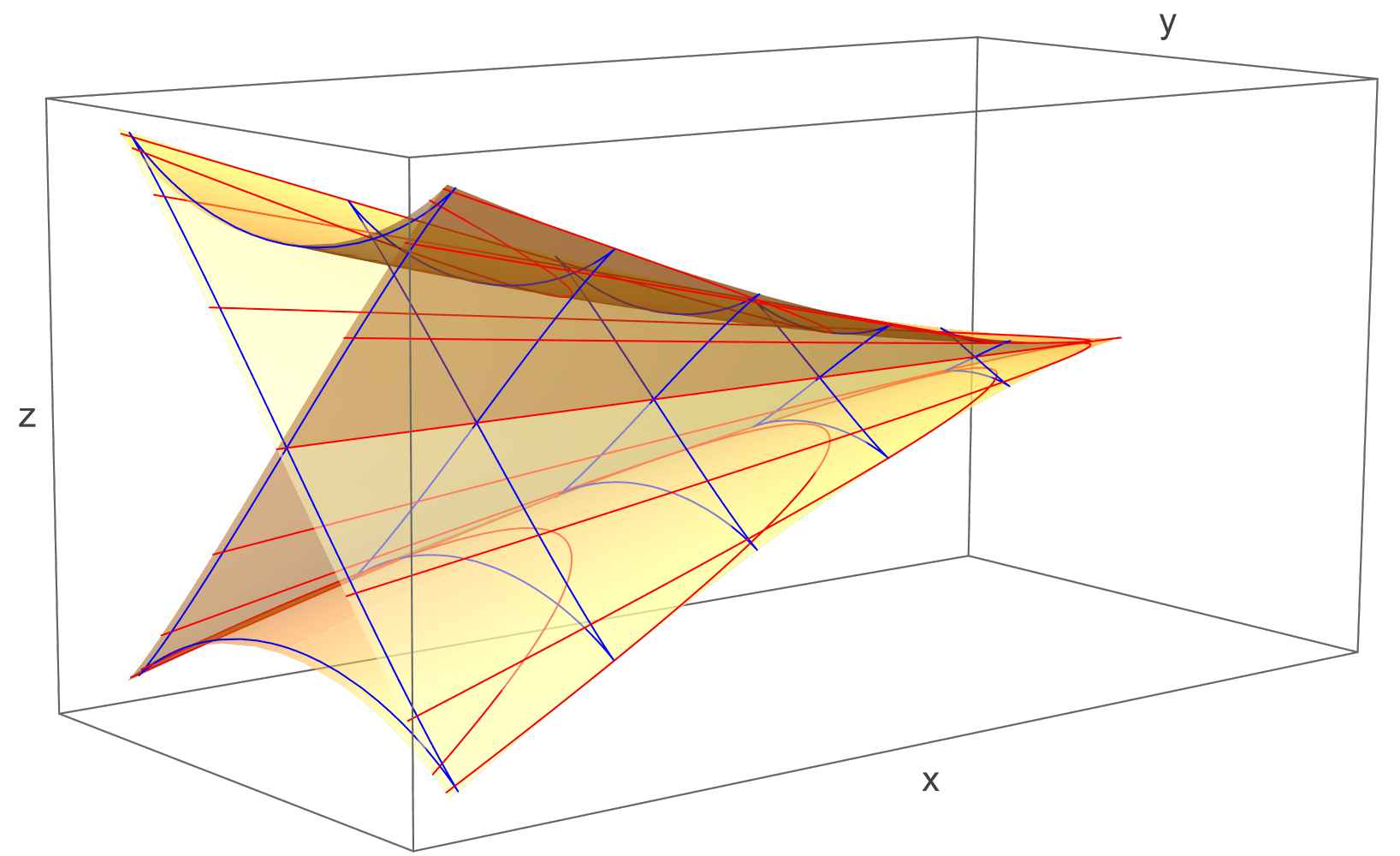}
    \caption{Local representation of a cuspidal vertex. The tip of the surface moves at the speed of light in the negative $z$ direction. Blue and red lines correspond to constant $z$ and $\sigma_1$ sections respectively.
    }
    \label{fig:cuspidal-vertex-representation}
\end{figure}

Note that at the moment of the vertex singularity, $\tau=0$, the equation defining the cuspidal vertex shape can be written, after rescaling coordinates, 
as 
\begin{equation}\label{vertexEQ}
    (x^2 - y^2)^3 = (x^3+3x\,y^2 -z^2)^2~,
\end{equation}
see Fig.~\ref{fig:cuspidal-vertex-representation}. This surface is manifestly invariant under the anisotropic scaling $(x,y,z)\to (\lambda^2 x, \lambda^2 y, \lambda^3 z)$. 
From this it is easy to see that the edges of the cross at $x=$constant sections scale in $x$ (towards the tip) as $y\sim x$, $z\sim x^{3/2}$. It also follows that near the cuspidal edge at any given $x$ (where both terms in \eqref{vertexEQ} vanish), with certain coordinate values $(x,y(x)+\delta y,z(x)+\delta z)$, the shape of the edge follows $\delta y^3\sim (\delta \hat z)^2$ with $\delta \hat z$ a rotated coordinate. Thus, the cross section of the edge is the same as for the  usual cosmic string cusp, $\delta z \sim (\delta y)^{3/2}$.



\subsection{Generic Ellipsoid}

The above analysis was made assuming a local (parabolic) parametrization of an arbitrary surface, so the question remains whether both types of cuspidal singularities arise in more generic surfaces. 

As discussed above, the wall rapidly attains relativistic velocities during its evolution, so that the ray tracing approximation \eqref{eq:eikonal approximation eq} becomes applicable and the formation of cuspidal edges becomes inevitable. For sufficiently pronounced departures from spherical symmetry, these singularities are expected to develop prior to complete collapse. In addition, the resulting cuspidal edges should propagate along the worldvolume, as observed in the axially symmetric example displayed in Fig.~\ref{fig:NG evolution}. A qualitatively similar behaviour is expected for generic surfaces. In the following, we will demonstrate this explicitly by studying the collapse of an ellipsoid with three unequal semi-axes.

\begin{figure}[t]
    \centering
        \includegraphics[width=\linewidth]{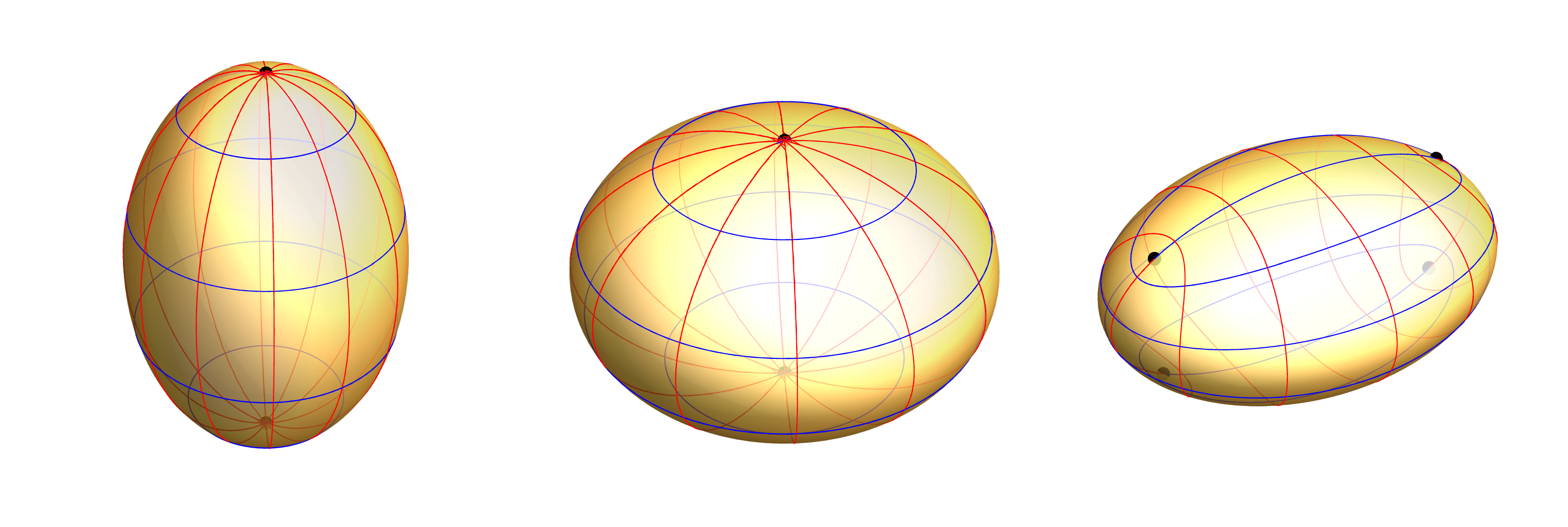}

    \caption{Oblate (left), prolate (middle), and triaxial (right) ellipsoids. The curves indicate lines of constant principal curvature. Umbilical points are shown as black dots.}
    \label{fig:types-of-ellipsoids}
\end{figure}

Concerning the cuspidal vertex singularity formation, it turns out that the same argument can be extended to a global property of any arbitrary (closed) surface. Indeed, in the eikonal approximation, the points of the wall can be understood as the wavefront of light rays. The set of points at which such rays generate caustics is called the focal set. For any given smooth surface embedded in Euclidean space $\mathbf{x}: \mathbb{R}^2\to \Sigma\in\mathbb{R}^3$ with unit normal field $\mathbf{n}$, the focal set is given by the two sets of points $\mathcal{F}_i(\Sigma)$ which satisfy
\begin{equation}
    \mathcal{F}_i=\qty{\mathbf{x}-\frac{\mathbf{n}}{\kappa_i}:\mathbf{x}\in\Sigma,\, (\kappa_{1,2}\text{ principal curvatures of $\Sigma$ at $\mathbf{x}$})}~.
\end{equation}

In general, the two sets $\mathcal{F}_i$ describe different surfaces, but they may coincide at the points $u$ where the surface $\Sigma$ is locally spherical, i.e., the principal curvatures coincide in value. Such points are called \emph{umbilical points}. Then, the focal points associated with the umbilic points, $\mathcal{F}_i(u)$, are precisely the ones at which the eikonal approximation predicts the largest focusing of energy, as the two focal surfaces merge at such points, i.e.
\begin{equation}
   u\text{  umbilical}\Leftrightarrow \mathcal{F}_i(u)\in \mathcal{F}_1(\Sigma)\cap \mathcal{F}_2(\Sigma)\Leftrightarrow \kappa_1(u)=\kappa_2(u).
\end{equation}
In particular, the singularities associated with the points $\mathcal{F}(u)$ belong precisely to the `umbilic family'~\cite{BERRY1980257}. 


\begin{figure}[t]
    \centering
        \includegraphics[width=1\linewidth]{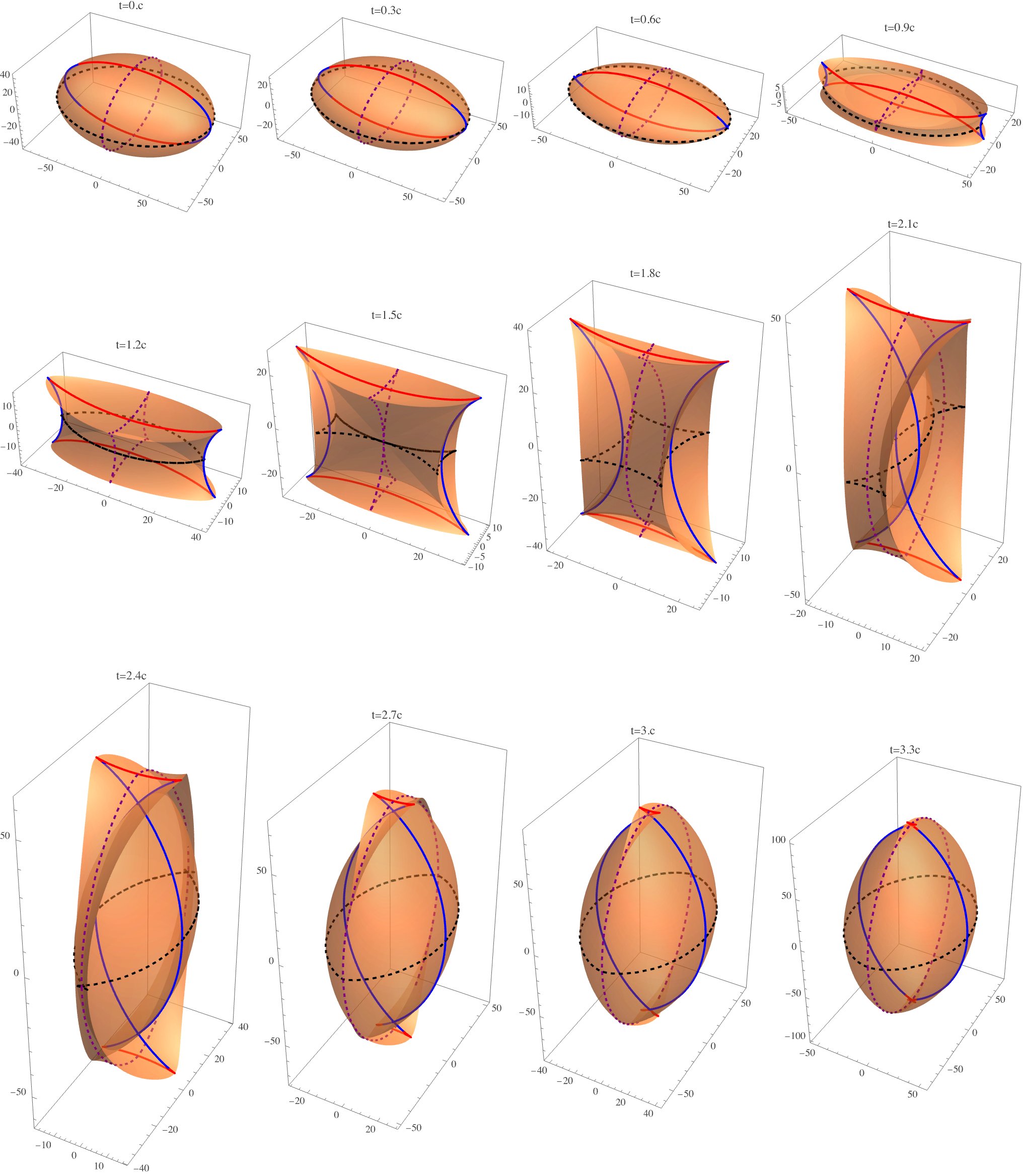}    
    \caption{
    Eikonal time evolution of a triaxial ellipsoid. As in Fig.~\ref{fig:eikonal_paraboloidA} cuspidal edges form (3rd-4th frame) and propagate along the worldsheet thereafter. They are not circular but have the characteristic unzipping lip structure seen in Fig.~\ref{fig:eikonal_paraboloidA}. Another swallowtail singularity forms on the lip (5th frame) that results into cuspidal edges moving across the lip. The cuspidal vertex (also seen in Fig.~\ref{fig:eikonal_paraboloidB}) corresponding to the umbilic points are seen in the 7th frame. According to \eqref{t_crit}, the cuspidal vertex formation time is $27/16\,c\simeq1.69 \,c$. 
    Three wall collision (self crossing) events occur at the center in frames 4, 6 and 8. 
    }
\label{fig:eikonal_triaxial}
\end{figure}

Within the ray tracing approximation, the question of the genericity of the cuspidal vertex may then be recast as the question of how generic umbilic points are on arbitrary convex closed surfaces in Euclidean space. Umbilic points on such surfaces typically arise as isolated points in convex regions and therefore generically form a discrete set. A well known conjecture due to Carath\'eodory states that any surface with the topology of a sphere must possess at least two umbilic points; see~\cite{GuilfoyleKlingenberg2008} and references therein. For generic initial conditions of closed domain walls, however, one generally expects a larger number of such points. As an illustrative example, the eikonal evolution of the triaxial ellipsoid shown in the rightmost panel of Fig.~\ref{fig:types-of-ellipsoids} is expected to contain four such hyperbolic umbilic points, and hence gives rise to four corresponding singular events. In the following we prove this statement by computing the eikonal development of generic triaxial ellipsoids.

Let us focus on the case of a generic triaxial ellipsoid with axes $a\leq b\leq c$,  defined implicitly by the equation
\begin{equation}
 \frac{x^2}{a^2} + \frac{y^2}{b^2}  + \frac{z^2}{c^2} = 1\,,  
\label{triaxial_config}
\end{equation}
or directly as the parametrization 
\begin{equation}
    \mathbf{x}_{\rm tri}(\theta,\varphi)=\mqty(a\sin\theta\cos\varphi\\
    b\sin\theta\sin\varphi\\
    c\cos\theta).   \label{parametric_triaxial}
\end{equation}

One may then show that this ellipsoid admits four umbilic points, given by~\cite{poelaert2011surface}
\begin{equation}
    \mathbf{u}=\qty{\pm a \sqrt{\frac{a^2 - b^2}{a^2 - c^2}},\,\, 0, \,\,\pm c\sqrt{\frac{b^2 - c^2}{a^2 - c^2}}},
    \label{umbilic_set}
\end{equation}
with normal vectors equal to\footnote{This is for  the umbilic point with only plus signs in \eqref{umbilic_set}, the others being obtained by reflections. }
\begin{equation}
    \mathbf{n}|_{\mathbf{u}}=\qty{\frac{c}{b}\sqrt{\frac{a^2-b^2}{a^2-c^2}},0,\frac{b}{c}\sqrt{\frac{b^2-c^2}{a^2-c^2}}}~.
\end{equation}

Using this parametrization of the umbilic points, one can derive the local form of the domain wall worldvolume in the vicinity of the singularity. The resulting expression is of the general form described above, with the coefficients in the expansion \eqref{local_umbilic} given by
\begin{equation}
    \kappa =\frac{ac}{b^3},\quad C=-3\frac{a c \sqrt{a^2-b^2} \sqrt{b^2-c^2}}{b^6}.
\end{equation}

Therefore, for a generic triaxial ellipsoid, the eikonal evolution generates four point-like singularities, located precisely at the intersections of the focal surface with the normals that pass through the umbilical points of the initial surface. In other words, a hyperbolic umbilic singularity (a cuspidal vertex) develops along the normal through each umbilical point, at a time given by the corresponding curvature radius,
\begin{equation}
    t_*=\kappa^{-1}=\frac{b^3}{ac}.
    \label{t_crit}
\end{equation}
Snapshots of this evolution are shown in Fig.~\ref{fig:eikonal_triaxial}.

For the special case of a prolate ellipsoid of revolution, in which $a=b$, the set of umbilic points degenerates to two, at the intersections of the wall with its symmetry axis. As we have seen, the singularities in this case form precisely on two points in this axis, at an instant given by $t_*=a^2/c\equiv a\sqrt{1-\epsilon^2}$, which coincides with our previous derivation. 

\subsection{Global results}

Moving towards more generic closed surfaces, it is natural to ask how generic the formation of cuspidal singularities is starting from smooth randomly perturbed closed walls. As argued above, cuspidal edges are expected to be generic for non-spherical initial shapes since points with one of the two principal radii of curvature smaller than the overall size are unavoidable, and these lead to swallowtail events that result into cuspidal edges.

Concerning cuspidal vertices, we have seen that in the eikonal  approximation (valid for thin wall in relativistic motion), each cuspidal vertex descends from an umbilical point in the smooth `initial' surface. The relevant question then is the population of umbilical points that can be expected. A result in differential geometry conjectured by Carath\'eodory states that any topological sphere must possess at least two umbilic points, see~\cite{GuilfoyleKlingenberg2008} and references therein. However, a larger number of them are expected to exist for generic initial conditions of closed walls. Indeed, for the triaxial ellipsoid shown in the center figure in Fig.~\ref{fig:types-of-ellipsoids} and whose eikonal evolution is shown in Fig.~\ref{fig:eikonal_triaxial}, there are four such points/events.

An estimate of the number of umbilic points per unit area on Gaussian random surfaces was presented in~\cite{Berry1977UmbilicPO}, where the surface density of umbilics was related to the moments of the power spectrum of fluctuations around a planar surface. Taking this result as indicative of the behaviour expected for more realistic domain wall configurations, one is naturally led to the conclusion that the formation of cuspidal vertices on domain walls should be a rather generic phenomenon, provided the conditions for eikonal evolution (relativistic and sufficiently thin walls) are met  throughout the collapse.  \\
How about the possibility of higher order singularities? In the case of wave fronts (i.e., in the eikonal approximation), a theorem by Arnold \cite{arnold1985} states that in three-dimensional space, wave fronts may have only self-intersections, cuspidal edges $(A_2)$ and swallow tails $(A_4)$; during propagation, singularities move along a caustic, and at certain moments of time rearrangements of the umbilical type ---for instance, cuspidal vertices $(D_4^+)$--- occur. This suggests  that the same is true for the NG propagation of a domain wall.

The next question is how generic these singularities are for completely general (closed) initial shapes, even if  not smooth. This case is of interest since the closed walls produced during the annihilation phase of a DW network are expected to have experienced  intersections and collisions with other walls, and this can result in kinks (or cuspidal points) in the `initial' shape. For cosmic strings in 3 dimensions it is well known that the presence of kinks in non-self intersecting loops can  avoid the cusp formation along the loop oscillation \cite{Garfinkle:1987yw}. 
The formation of cuspidal singularities at later stages of the evolution may be compromised by the presence of 
non-smooth features in the initial collapse configuration.
Hence, it is fair to address whether singularities should be expected generally.

The key to answer this more general question comes from a powerful insight from differential topology: the fascinating problem of the sphere eversion \cite{smale},  see \cite{FrancisSullivan1998,Bednorz_2019} for introductions. In topology, a smooth eversion is a continuous deformation where the surface is turned inside-out by allowing  self-intersections but no singular points (with degenerate induced metric) in the process. A second look at Figs.~\ref{fig:NG evolution} and \ref{fig:eikonal_triaxial} reveals that indeed the DW motions  precisely realize eversions (of singular type). This is particularly evident from the eikonal limit: in the last panels of Fig.~\ref{fig:eikonal_triaxial} each point comes from near its antipodal pair, so globally the DW is `inside out'. 
Hence,  contracting DWs execute physical eversions. In the examples above both of them are singular, they do not qualify as proper mathematical smooth eversions -- and for a good reason. 


The topological problem of sphere eversion is particularly useful because it is independent of the specific dynamical law governing the wall. Instead, it allows for arbitrary continuous deformations and therefore isolates the purely kinematical constraints imposed by the requirement that the surface evolve without punctures or tears. In this setting, self-intersections are permitted—and, in fact, are necessary for the eversion to occur. This is directly relevant to domain wall dynamics. Although the detailed evolution of a domain wall depends on the microscopic model—for example, on whether the motion is governed purely by the NG equations, modified by additional forces on the worldvolume, or affected by other microscopic degrees of freedom—the basic kinematical rules are analogous: the wall cannot puncture or terminate, while self-intersections are allowed. Consequently, the sphere eversion problem provides a useful topological guide to the types of singular structures that may arise in domain-wall collapse beyond the specific assumptions of the NG approximation.

Famously, mathematical smooth eversions do exist \cite{smale}. Heuristically, the crucial ingredient is that sufficient  negative curvature allows the eversion with self intersections only. 
%
It will suffice for the scope of this work to mention a more concrete result: that  smooth eversions require that in some  intermediate stage the surface has at least one quadruple point (four sheets of the surface passing through it) \cite{MaxBanchoffEversion,Hughes1985,MinimaxEversion}.
Hence, topology alone does not completely close the door to collapsing DWs without any cuspidal singularities.


However, it clarifies the enormous challenge: the surface must develop 
negative curvature and a quadruple point. In plain words, intermediate shapes are significantly intricate, see e.g. in  \cite{FrancisSullivan1998,Bednorz_2019}. We cannot offer here a proof but it appears very unlikely that  such  magnificent sequence of surfaces can be the outcome of the dynamics governing DW collapse, even allowing complete freedom in the initial velocity field. 
Still, the topological argument seems strong enough to establish that even if there were initial conditions leading to such smooth eversion they would be strongly fine tuned. In other words, with generic initial conditions singularities are unavoidable. 

We can, then, safely conclude that generically physical closed DWs are expected to form cuspidal singularities. Provided the DW is thin enough, these should be realized. The eversion argument does not inform on whether they are cuspidal edges or vertices, however, the former are clearly more generic.

\section{Cuspidal Singularities in Field Theory}  
\label{Sec:FT}

In the previous sections, we have discussed the formation of swallowtail singularities, cuspidal edges, and cuspidal vertices in the dynamical evolution of non-spherical membranes, both in the Nambu--Goto framework and in the eikonal (wavefront) approximation. However, within field theory, domain walls are not infinitely thin membranes, but solitonic objects of finite thickness and nontrivial internal structure, arising generically in nonlinear theories with discrete sets of vacua. The descriptions employed above are therefore to be understood as effective, low energy approximations to the dynamics of such walls, valid whenever the details of their internal structure do not play a significant role.

It is then natural to ask whether the singular structures identified above might cease to form once the relevant geometric scales become comparable to the wall thickness in its rest frame. The main aim of this section is to show that this is not the case, at least at the qualitative level: swallowtail singularities, cuspidal edges, and cuspidal vertex singularities do indeed form in the full field theory, although their profiles are, as expected, smoothed by the emission of some radiation from these regions due to the finite width of the wall. This is analogous to the emergence of cusps in field theory simulations of cosmic strings~\cite{Olum:1998ag}. This suggests that thin wall descriptions retain their predictive validity beyond the regime in which they might, on naive grounds, be expected to apply.

\subsection{Simulation setup}

We consider a simple scalar field model with a $\lambda\phi^4$ potential in four-dimensional flat spacetime, with action
    \begin{equation}
        S=\int d^4 x \left[\frac{1}{2} \eta^{\mu \nu} \partial_\mu \phi \partial_\nu \phi-\frac{\lambda}{4}\left(\phi^2-\eta^2\right)^2\right],
    \label{FT_action}
    \end{equation}
where $\eta$ and $\lambda$ are arbitrary positive constants, the former having dimensions of energy and the latter being dimensionless. The action \eqref{FT_action} allows for static solutions given by flat, infinite domain walls for which the profile of the scalar field along the codimension interpolates between the two disconnected points of the vacuum manifold following the one-dimensional \emph{kink} solution:
\begin{equation}
\label{kink-f4}
\phi_K(x) = \eta \tanh \left[\delta^{-1} (x - X)\right]\,,
\end{equation}
where $X$ represents the position of the soliton along the codimension, and 
$\delta=\sqrt{\frac{2}{\lambda} }\left(\frac{1}{\eta}\right)$ is the 
characteristic width of the soliton profile.

We initiate our field theory simulations with an ellipsoidal scalar field configuration given by
\begin{equation}
    \phi_{\rm init}=\eta \tanh\qty(\frac{d}{\delta}) \,,
\end{equation}

\noindent where $d$ is the signed perpendicular distance from the ellipsoid given by \eqref{triaxial_config}, with the outward direction being positive. We choose $\lambda = 0.5$ and $\eta = 1.0$, so in our units $\delta = 2.0$. In all our simulations, we will assume that the radius of curvature at every point of the initial domain wall is much larger than the wall's characteristic width, so that higher curvature corrections to the field profile can be neglected.

We use \textsc{GRChombo}~\cite{Andrade:2021rbd}, a multipurpose numerical relativity code that has the advantage of an in-built adaptive mesh refinement (AMR) system that allows us to evolve high-resolution simulations at lower computational cost than fixed resolution codes\footnote{Note, however, that for the purposes of this paper we have not used the capabilities of \textsc{GRChombo} of evolving the gravitational field. In fact we have performed all the field theory simulations in flat spacetime.} . For the case without axisymmetry we use \textsc{GRDzhadzha}~\cite{Aurrekoetxea:2023fhl}, a modified version of GRChombo in which the background spacetime is fixed. This allows us to run full 3D simulations at minimal computational cost. Our computational domain has side length $L = 128 $ and coarse resolution $dx_c = 2.0 $ with the Sommerfeld (radiative) outer boundary condition~\cite{Alcubierre:2002kk}. Though this is only valid for massless radiation, the boundary is sufficiently far from the collapsing wall as to not noticeably affect our results. We use nine levels of refinement, with each level having double the resolution of the previous one, such that the finest resolution has $dx_f = 0.004$ (to 3 d.p). The AMR is set so that it tracks regions with the highest energy density (i.e. the domain wall) and adds higher levels of refinement to regions in which the energy density increases above a certain chosen threshold. This tends to progressively occur as the wall collapses, due to its increasing kinetic and gradient energy as it accelerates and becomes more Lorentz contracted. Thus initially the wall is covered by five levels, with more being added as the collapse progresses. 
\begin{table}[h!]
\begin{center}
\begin{tabular}{||c |c c c ||} 
 \hline
 Shape & $a$ & $b$ & $c$ \\ 
 \hline\hline
 Prolate spheroid & 80 & 40 & 40 \\ 
 \hline
Oblate spheroid & 80 & 40 & 80 \\
 \hline
  Triaxial ellipsoid  & 80 & 60 & 40 \\
\hline
\end{tabular}
\caption{Parameters of the triaxial ellipsoid equation \eqref{triaxial_config} for each configuration simulated.  Note that with our units the wall has a thickness $\delta = 2.0$.}
\end{center}
\end{table}
For spheroidal domain wall configurations, we take advantage of the axial symmetry to reduce the computational cost of our simulations. Usually, one would do this by adopting cylindrical coordinates; however, due to the use of AMR, we are restricted to a Cartesian system. Fortunately, one can still harness this symmetry by applying the \textit{cartoon method} (see~\cite{Cook:2016soy}), which allows one to use AMR in a 2D cartesian domain corresponding to the $r-z$ plane in cylindrical coordinates ($z$ being the axis of rotation). On the other hand, for the triaxial system we were forced to evolve an octant of a 3D cartesian grid, invoking reflection symmetry across two planes. To be sure of the validity of our results, we perform a point-wise convergence test for the prolate spheroid case along the rotation axis through the swallowtail feature (see \cref{Ap:2}).

\subsection{Results}

\begin{figure}[h!]
    \centering
    \includegraphics[width=1.0\linewidth]{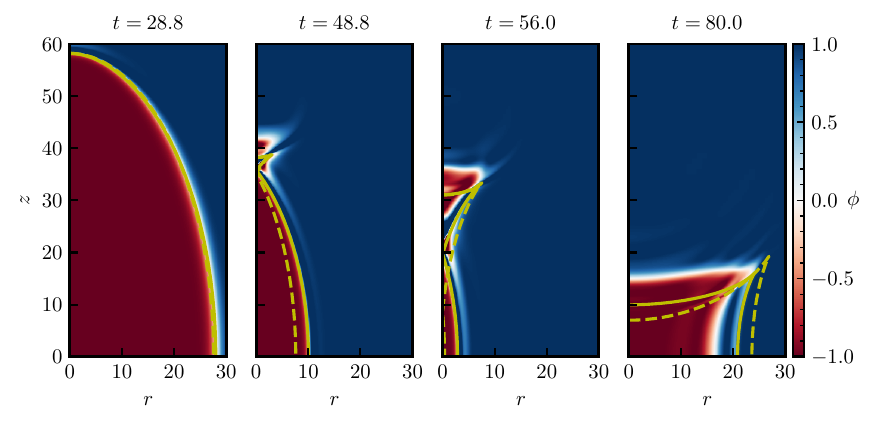}
    \caption{The scalar field at different times according to the field theory evolution for a prolate spheroid (the field is rotated about the $z$-axis and reflected across the $r$-axis). We illustrate the Nambu-Goto and eikonal approximations with continuous and dashed yellow lines, respectively. One observes the formation of swallowtail-like structures and the expansion of the subsequent cuspidal edges in the second and third plots, with the final plot showing the configuration shortly after the axisymmetric collapse along the z-axis. Remarkably, this swallowtail structure matches fairly well with the approximations. }
    \label{fig:prolate_compare}
\end{figure}
\begin{figure}[h!]
    \centering
    \includegraphics[width=1.0\linewidth]{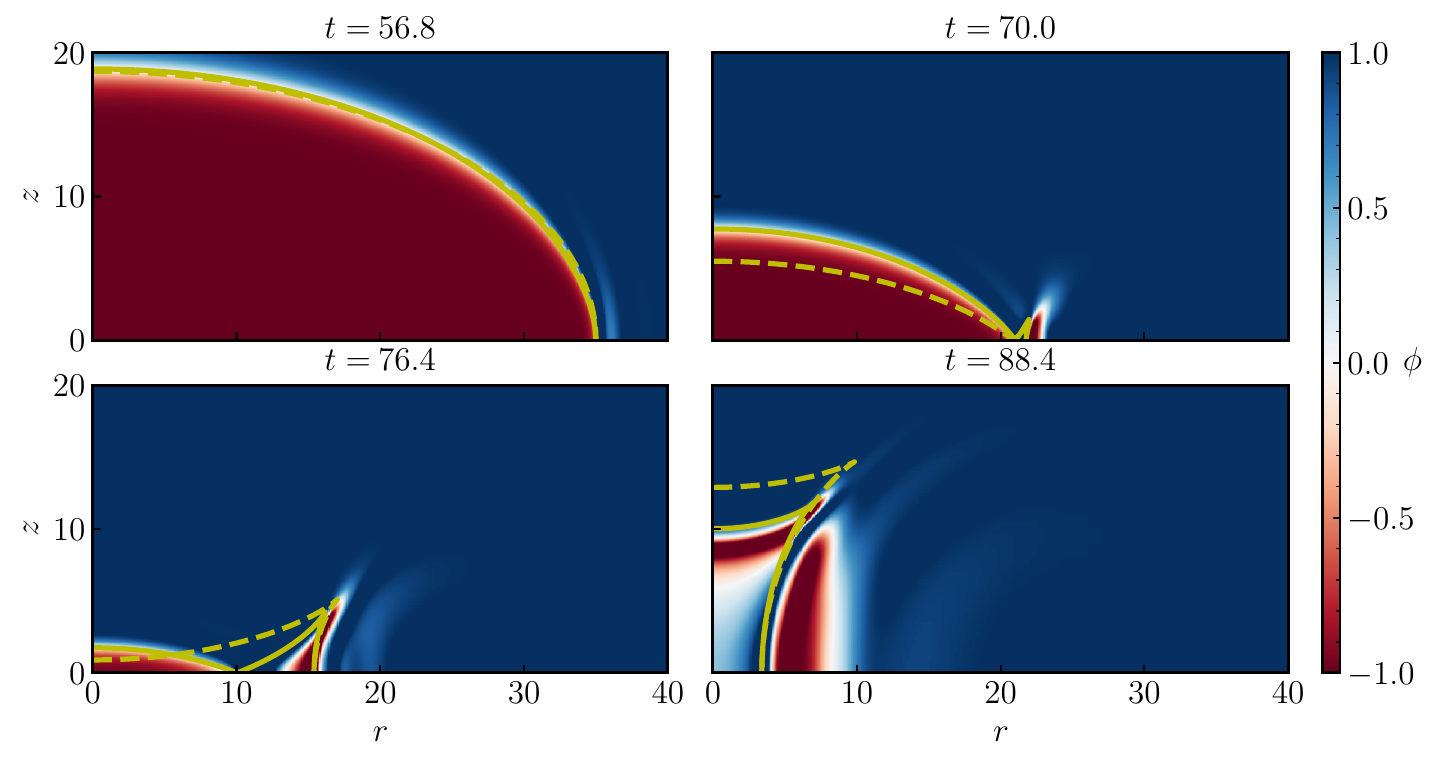}
    \caption{The same as Fig.~\ref{fig:prolate_compare} but for the oblate case. As the field is rotated about the $z$-axis, the ``swallowtail'' structure here extends across the perimeter of the configuration, and thus the focusing of the field is less significant. }
    \label{fig:oblate_compare}
\end{figure}

In Fig.~\ref{fig:prolate_compare} and Fig.~\ref{fig:oblate_compare}, we illustrate the evolution of the field at different times for the prolate and oblate cases, respectively. We also show the trajectories of the NG (yellow solid line) and eikonal (yellow dashed line) approximations. It should be noted that we only compare with the eikonal approximation once the regions of highest initial curvature have reached a sufficiently high velocity ($v = 0.99$), given that the approximation holds at constant (light) speed. One observes that for both cases the field theoretical evolution is remarkably similar to both the NG and eikonal predictions. Naturally, the NG trajectory tracks the wall more accurately than the eikonal approximation. The latter incorrectly predicts collapse in a shorter time, as the part of the wall which initially has lower curvature has a smaller acceleration and thus has not quite reached a constant velocity by the time the eikonal approximation is applied. Thus, one expects an improvement in the eikonal approximation when tracking the collapse of larger domain wall configurations, as one can then initiate the eikonal approximation at later times when all parts of the wall have reached the maximum velocity.

One observes that the domain wall develops structures akin to the swallowtail predicted by both approximations (at $t = 48.8$ and $t = 70.0$ in Fig.~\ref{fig:prolate_compare} and Fig.~\ref{fig:oblate_compare} respectively). Moreover, smoothened cuspidal-like structures emerge from this and expand outwards. These are presumably the field theory analogues of the cuspidal edges seen in the approximations described earlier. For the prolate case, the NG solution does not quite align with the base of the swallowtail structure (see third panel of Fig.~\ref{fig:prolate_compare}). This is arguably a consequence of the high degree of focusing when the swallowtail forms, during which NG will naturally lose its validity. This line of reasoning is supported by looking at the oblate case, for which the NG trajectory more closely matches the swallowtail profile. For this case, the focusing is less extreme, as it occurs all along a circle - the perimeter of the configuration - rather than at a point, and so NG appears to retain better accuracy.

\begin{figure}[h!]
\centering
 \includegraphics[width=0.43\linewidth]{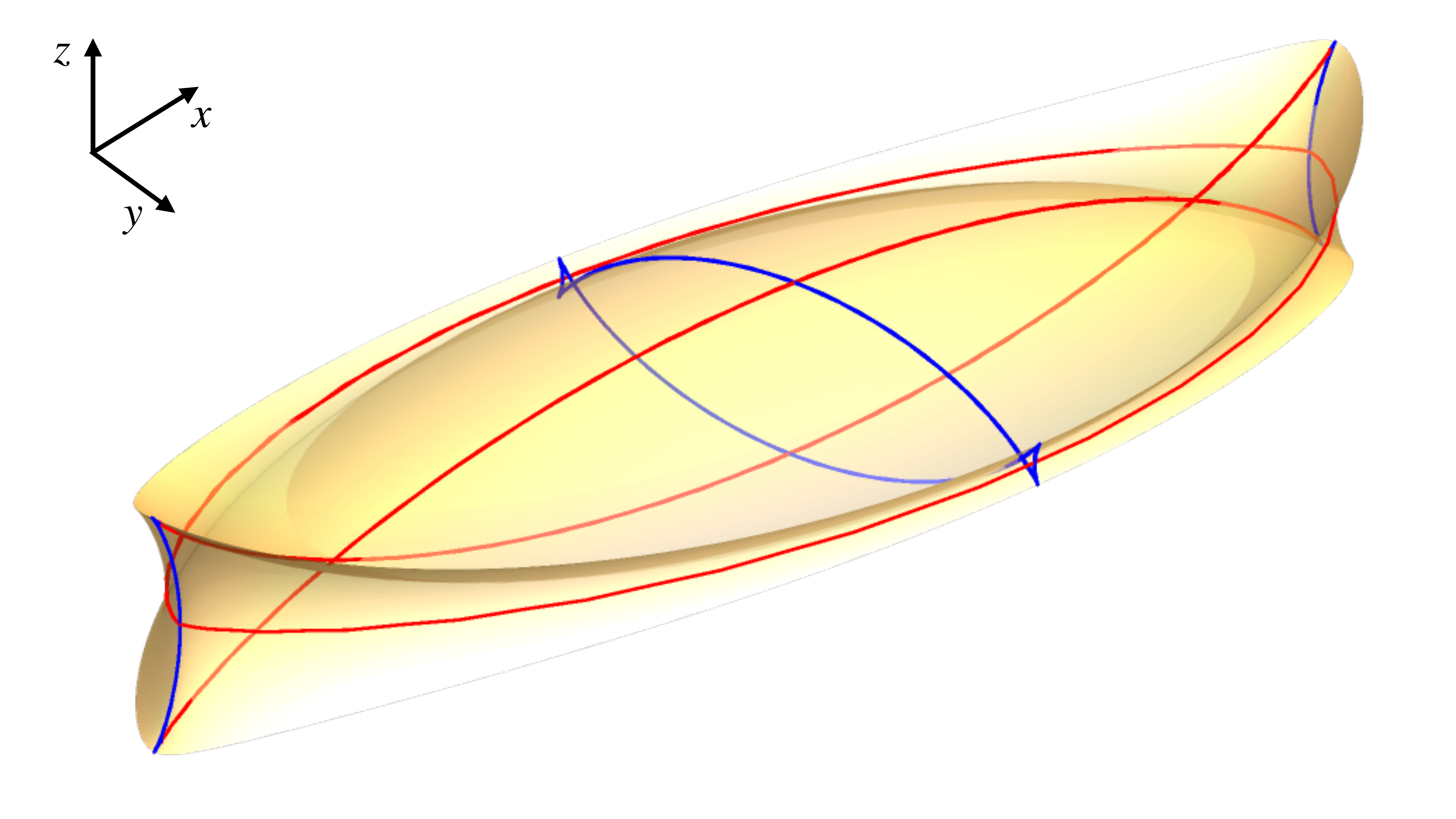}
\includegraphics[width=0.36\linewidth]{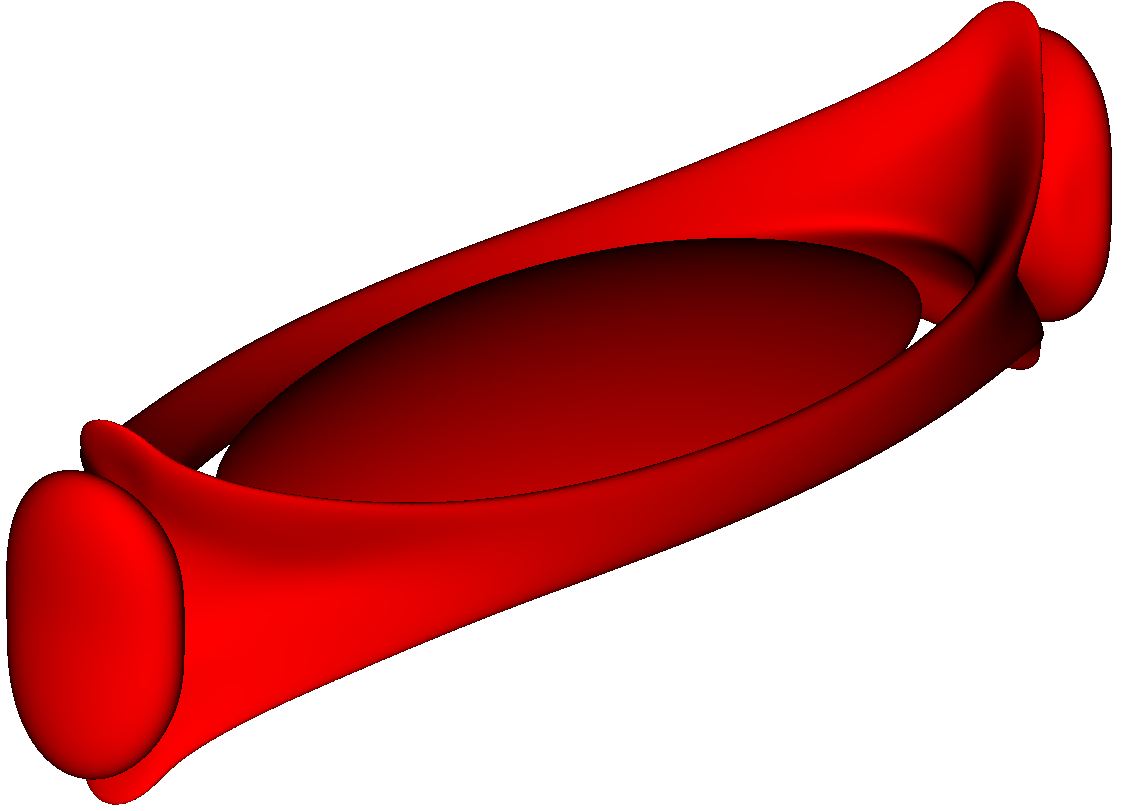}
\includegraphics[width=1.0\linewidth]{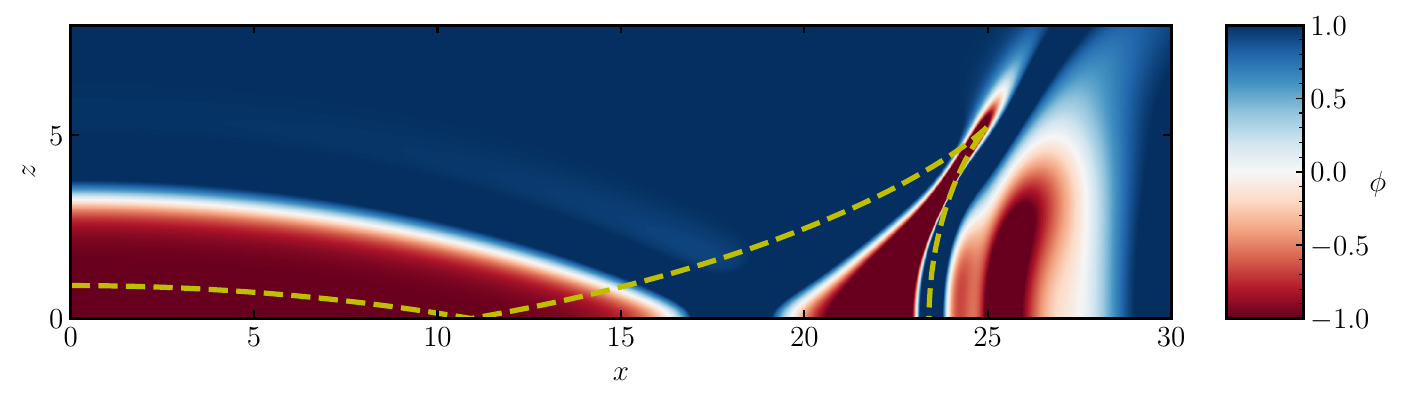}
\caption{Upper: 3D snapshot of the collapsing triaxial ellipsoid domain wall using the eikonal approximation (left) and field theory for $\phi = 0$ (right). One can see the cuspidal edges around the perimeter of the ellipsoid. More importantly, lip-shaped formations grow at each of the four tips of the configuration. These are the smoothened field theory version of the cuspidal vertices predicted by both approximations. Lower: cross-section of the field in the $y = 0$ plane, which cuts along the centre of the object and through the cuspidal vertex.}
\label{fig:triaxial_compare}
\end{figure}

Furthermore, at the top of Fig.~\ref{fig:triaxial_compare} we illustrate a 3D snapshot of the triaxial ellipsoid collapse in the eikonal approximation (left) and field theory simulation (right), with a longitudinal cross-section shown below. Again, one observes structures present in the axially symmetric cases - the swallowtail and cuspidal edges - as well as the four cuspidal edges converging at the tips to form smoothened cuspidal vertices. Thus, it appears that the validity of the approximation is not just restricted to axially symmetric phenomena. We therefore conclude that the formation of cuspidal singularities as predicted in the thin wall approximations does in fact reproduce quite accurately the true behavior of the wall in field theory, and hence they are not artifacts produced by the failure of such approximations.

Singularities in the wall worldvolume must correspond to large values of field gradients in the field theoretical evolution and subsequently to a local focusing of energy. We can compute the energy density of the field in the laboratory frame at every instant of time, which is simply given by
\begin{equation}
    \rho \equiv T_{00}(\mathbf{x},t)= \frac{1}{2}(\partial_0\phi)^2+\frac 1 2\partial_i\phi\partial^i\phi+ V(\phi)~.
\end{equation}
Thus we expect that the maximum energy density $\rho_{max}$ is localised in the neighborhood of the singularity formation event. In particular, the time of singularity formation is predicted from the eikonal approximation to be the inverse of the curvature radius at the umbilic points. In the case of a triaxial ellipsoid, it can be found analytically in terms of the corresponding semi-axes (see \cref{t_crit}). Crucially, in optics higher-order catastrophes are known to yield a greater degree of focusing of intensity when the wave nature of light is taken into account, using the stationary‑phase expansion of a diffraction integral \cite{BERRY1980257,Jaramillo:2022mkh}. Thus one expects the global maximum of the energy density to occur wherever the approximations predict the highest-order catastrophe forms: for the axially symmetric case it is the swallowtail, and for the triaxial case it the cuspidal vertex.

\begin{figure}[h!]
    \centering
    \includegraphics[width=0.9\linewidth]{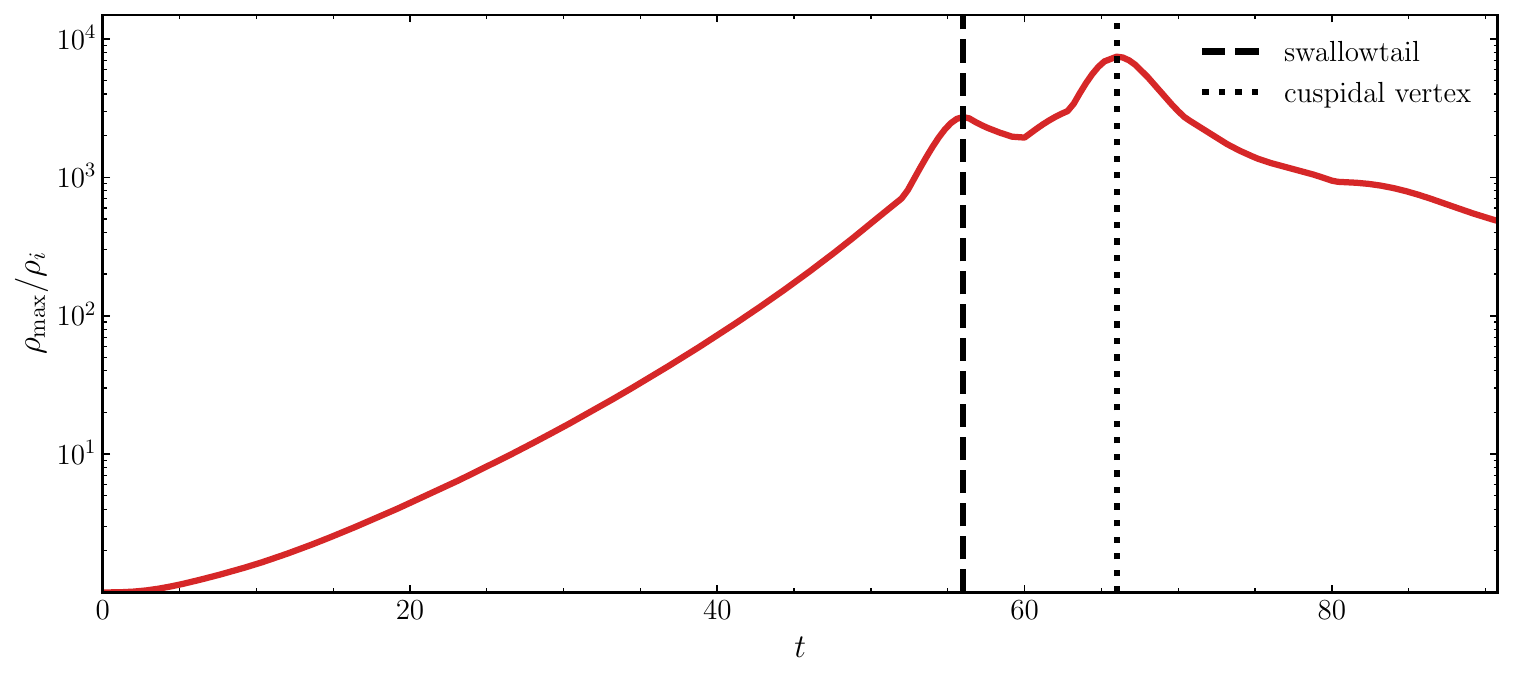}
    \caption{Time evolution of the global maximum of the energy density $\rho_{\mathrm{max}}$ (normalised by the initial energy density at the centre of the wall $\rho_i$) for the collapse of the triaxial ellipsoid. The two peaks correspond to the formation of the swallowtail followed by the cuspidal vertices. One observes that the cuspidal vertices exhibits the highest energy concentration as a consequence of being a high-order catastrophe.}
    \label{fig:rho_max}
\end{figure}

\begin{figure}[h!]
    \centering
    \includegraphics[width=0.9\linewidth]{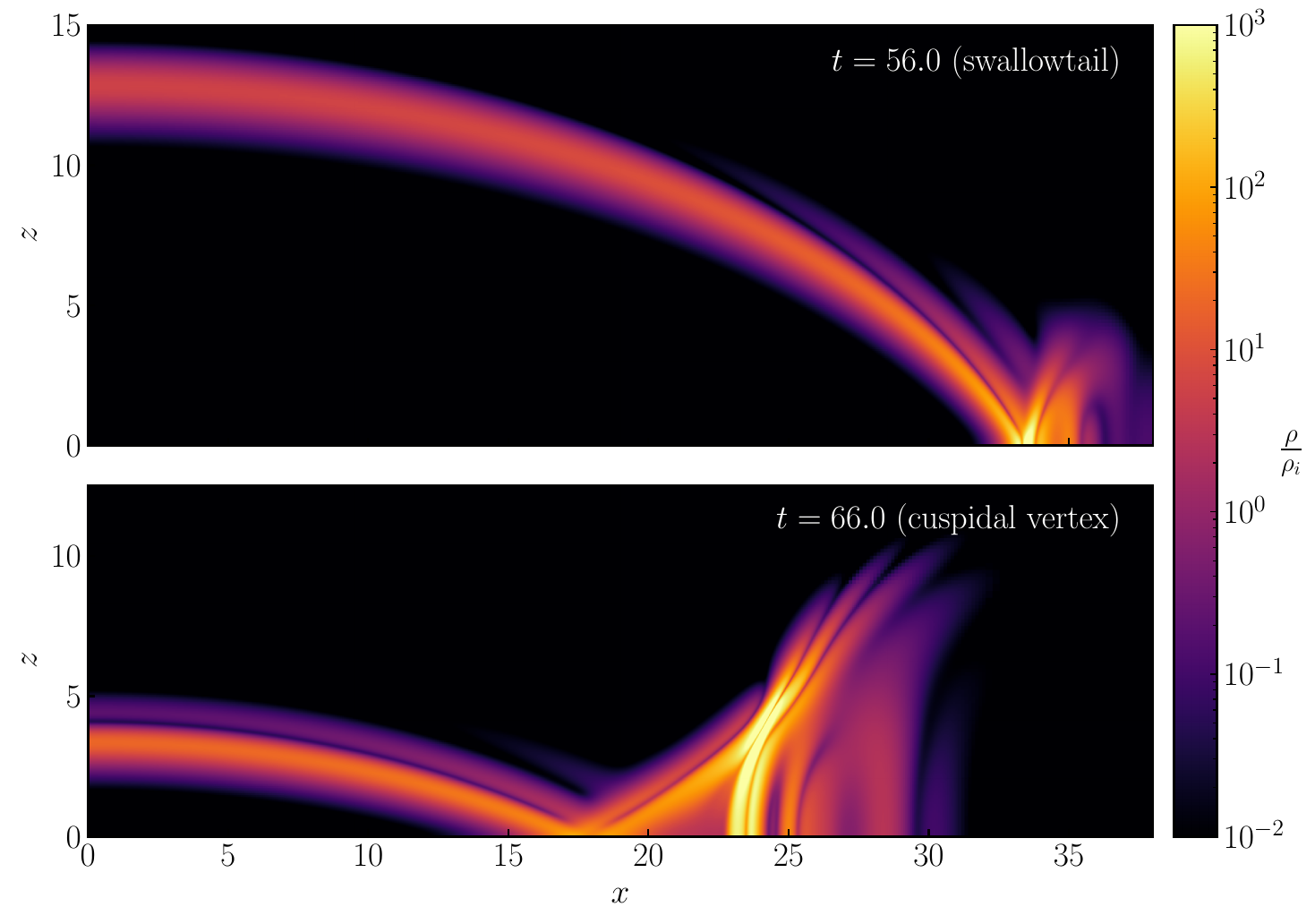}
    \caption{Cross-sections of the energy density (normalised by the initial energy density at the centre of the wall $\rho_i$) in the $y = 0$ plane for the triaxial ellipsoid collapse. Upper: swallowtail formation at $t = 56.0$. Lower: cuspidal vertex development at $t = 66.0$ - this is the highest the energy density reaches.}
    \label{fig:triaxial_rho_plot}
\end{figure}

This expectation is shown to be valid by Fig.~\ref{fig:rho_max}, which illustrates the evolution of the ratio of the global maximum of the energy density to its initial value (at the wall's center). One observes a steady growth in the maximum energy density, which is located at the poles of the ellipsoid. This growth is a consequence of the increase in the kinetic energy experienced by the entire domain wall as well as the predicted localised focusing around the poles. There is a subsequent peak in $\rho_{\mathrm{max}}$ at the swallowtail location at $t = 56.0$, followed by a slight decrease as the cuspidal edges expand outwards. However, this is short-lived as the cuspidal vertex forms at $t = 66.0$ to yield the highest energy concentration. Both of these events can be seen in the energy density cross-section plots shown in Fig.~\ref{fig:triaxial_rho_plot}.

If the energy density concentrated in these focusing events is sufficiently high in their local rest frame, the resulting localized gravitational collapse could, in principle, lead to black-hole formation. To fully explore this possibility one naturally requires the inclusion of backreaction, and so is beyond the scope of this work. We therefore leave this to future studies that invoke full numerical relativity. 

\section{Conclusions}
\label{Sec:Conclusions}

In this work, we have shown that the collapse of closed relativistic domain walls generically gives rise to worldvolume singularities in $3+1$ dimensions. More specifically, we have identified and characterized two distinct classes of singular structures: cuspidal edges, namely one dimensional singular loci that propagate along the wall for a finite interval of time, and cuspidal vertices, which are higher order, localized, and instantaneous singular events. A key feature of both classes is that the wall reaches relativistic velocities in the local vicinity of the singular region.  In this sense, these events constitute the natural domain wall analogue of the cusps known to arise in cosmic string evolution~\cite{Turok:1984cn}. We have further demonstrated that these structures arise dynamically from smooth, non-spherical initial conditions and therefore represent a robust outcome of relativistic domain wall evolution, rather than an exceptional consequence of finely tuned configurations. Their formation is ultimately driven by the nonlinear collapse of closed walls under their own tension, which rapidly pushes the system into the ultra-relativistic regime.

Our analysis shows that the formation and evolution of these singularities can be understood within a thin wall description. 
Moreover, these events admit a clear characterization within catastrophe theory: cuspidal edges are associated with swallowtail-type bifurcations, whereas cuspidal vertices correspond to higher order focusing events in catastrophe theory. We have also shown that, once the wall becomes ultra-relativistic, an eikonal, or ray tracing, approximation provides an accurate description of the global evolution and correctly reproduces the singularity structure. This makes the genericity of the phenomenon especially transparent, since it follows directly from the focusing properties of the inward normal flow.

A comment is in order regarding the potential impact of worldsheet curvature corrections on deviations from pure Nambu–Goto (NG) dynamics. As shown recently by some of the authors \cite{Blanco-Pillado:2024bev}, the effective action governing the motion of $3+1$-dimensional domain walls generally contains curvature dependent contributions, in particular a term proportional to the Ricci scalar of the worldsheet. Since the Ricci scalar diverges as the worldsheet approaches a cuspidal singularity, one might expect such terms to become dominant in the vicinity of the forming cusp and thereby alter, or even prevent, the development of the singularity.
This expectation is, however, misleading. In the case of relativistic strings, curvature corrections of this type are absent from the effective action \cite{Aurrekoetxea:2026xrl}, and therefore the NG description remains valid all the way to cusp formation \cite{Olum:1998ag,Blanco-Pillado:2025gzs}. For domain walls, although curvature corrections are present, these corrections are not strong enough to counteract the NG dynamics responsible for the formation of cuspidal singularities. The emergence of such singularities in the evolution of $3+1$-dimensional domain walls is therefore robust and cannot be prevented by these higher‑curvature effects in the worldvolume effective action.

A central result of this paper is that these singular structures are not artifacts of the thin wall approximation. Our adaptive mesh refinement field theory simulations reproduce, at the qualitative level, the same cuspidal edges, cuspidal vertices, and associated bifurcation patterns predicted by the thin wall analyses. Although the finite thickness of the wall smooths the singular profiles somewhat, the underlying focusing events remain clearly visible and occur at the locations, and with the morphologies, anticipated from the effective description. This agreement strongly supports the interpretation of these features as genuine aspects of the nonlinear dynamics of solitonic domain walls.

The field theory evolution further shows that these catastrophes are accompanied by strong localization of energy density and curvature on the wall worldvolume, leading to localized regions of extreme field gradients. This suggests that cuspidal singularities may play an important role in the microscopic dynamics of collapsing domain wall networks and may act as localized sources of enhanced radiation. In particular, they provide a plausible mechanism for generating ultraviolet structure in the gravitational wave signal emitted during domain wall annihilation, beyond what is captured by estimates based solely on the macroscopic scale of the network. They may therefore offer a natural explanation for the excess high frequency power hinted at in recent numerical studies of gravitational-wave production from domain walls~\cite{Notari:2025kqq}. Since this phenomenon is intrinsically ultraviolet in character, it may not be fully resolved in simulations with coarse lattice spacing, thereby motivating a careful reassessment of existing network simulations from this perspective.

Finally, we expect the conclusions of this work to extend beyond the specific case of pure NG dynamics. The formation of cuspidal edges and cuspidal vertices is ultimately controlled by the kinematics of a collapsing closed surface and by the focusing properties of the associated normal flow. These ingredients are not tied to the precise form of the NG equations. In particular, similar singular structures should also arise in more general domain wall dynamics, including situations in which the wall evolves under the action of additional forces. A physically relevant example is provided by biased domain walls separating vacua with different energy densities, for which the corresponding pressure difference accelerates the wall in addition to its intrinsic tension. While such effects can modify the quantitative details of the collapse, including the time of formation, location, and local coefficients characterizing the singular event, they are not expected to change the local catastrophe type of the generic focusing structures. From this perspective, the singularities identified here should be regarded as robust features of relativistic domain wall collapse, rather than as special consequences of the idealized NG limit.

These results open several directions for future work. A first priority is to quantify the gravitational wave emission from individual focusing events and to assess their cumulative contribution to the spectrum produced by collapsing domain wall networks. Another important question concerns the regime of validity of the flat space approximation. Although gravitational backreaction is usually negligible for the overall evolution of domain walls, the large local energy densities attained near cuspidal singularities may invalidate this approximation in the immediate vicinity of these singular events. In such regions, strong gravity effects may become relevant affecting the local dynamics and could even trigger primordial black hole formation. This raises the intriguing possibility that unstable domain wall networks in the Early Universe may source not only gravitational radiation, but also primordial black holes through singularity driven energy focusing events.

Overall, our results indicate that cuspidal singularities are a generic and physically relevant component of domain wall collapse. They provide a new perspective on the nonlinear dynamics of relativistic domain walls, connect naturally with the mathematics of wavefront singularities and catastrophe theory, and point toward potentially important consequences for the cosmological signatures of unstable domain wall networks. A quantitative assessment of the connection between the singularity structures identified here and their observable implications lies beyond the scope of the present work and is left for future investigation.

\acknowledgments
We would like to thank Ken D. Olum and Ander Urtiaga Bermudez for helpful conversations. A.G.M.-C. thanks Jose Luis Jaramillo and Oscar Meneses-Rojas for invaluable discussions on catastrophe theory.
This work has been supported in part by the 
PID2024-156016NB-I00 grant funded by MICIU/AEI/ 10.13039/501100011033 by "ERDF/EU",
the Basque Government grant
(IT-1628-22) and the Basque Foundation for Science (IKERBASQUE). The work of A.G.M.-C. is also supported
by Grants No. ED481B-2025/059 and ED431B-2024/42 (Xunta de Galicia). D.~J.-A. is supported in part by National Science 
Foundation grant PHY-2419848.
O.P. and J.S.V.B. acknowledge support from the Spanish Ministry of Science and Innovation (MICINN) through the Spanish State Research Agency under the R\&D\&i project PID2023-146686NB-C31 funded by MICIU/AEI/10.13039/501100011033/ and by ERDF/EU, and under Severo Ochoa Centres of Excellence Programme 2025-2029 (CEX2024001442-S). IFAE is partially funded by the CERCA program of the Generalitat de Catalunya. ME acknowledges support of the European Structural and Investment Funds and the Czech Ministry of Education, Youth and Sports (project FORTE — CZ.02.01.01/00/22\_008/0004632). This work also used the DiRAC Memory Intensive service Cosma8 at Durham University, managed by the Institute for Computational Cosmology on behalf of the STFC DiRAC HPC Facility (\url{www.dirac.ac.uk}). The DiRAC service at Durham was funded by BEIS, UKRI and STFC capital funding, Durham University and STFC operations grants. DiRAC is part of the UKRI Digital Research Infrastructure.

\bibliography{biblio}

\newpage
\appendix

\section{Convergence test}
\label{Ap:2}

As a check of the validity of our field theory simulations, we perform a local convergence test through the swallowtail feature that develops during the prolate spheroid collapse. By comparing the scalar field profile along the rotation axis for runs with three base resolutions $dx_{high} = 2.0$,  $dx_{med} = 4.0$ and  $dx_{low} = 8.0$ (keeping 9 refinement levels) one can identify the order of convergence

\begin{equation}
p = \log_2 \frac{\left|\phi_{\text {low }}-\phi_{\text {med}}\right|}{\left|\phi_{\text {med }}-\phi_{\text {high }}\right|} \,.
\end{equation}

\begin{figure}[h!]
    \centering
    \includegraphics[width=1.0\linewidth]{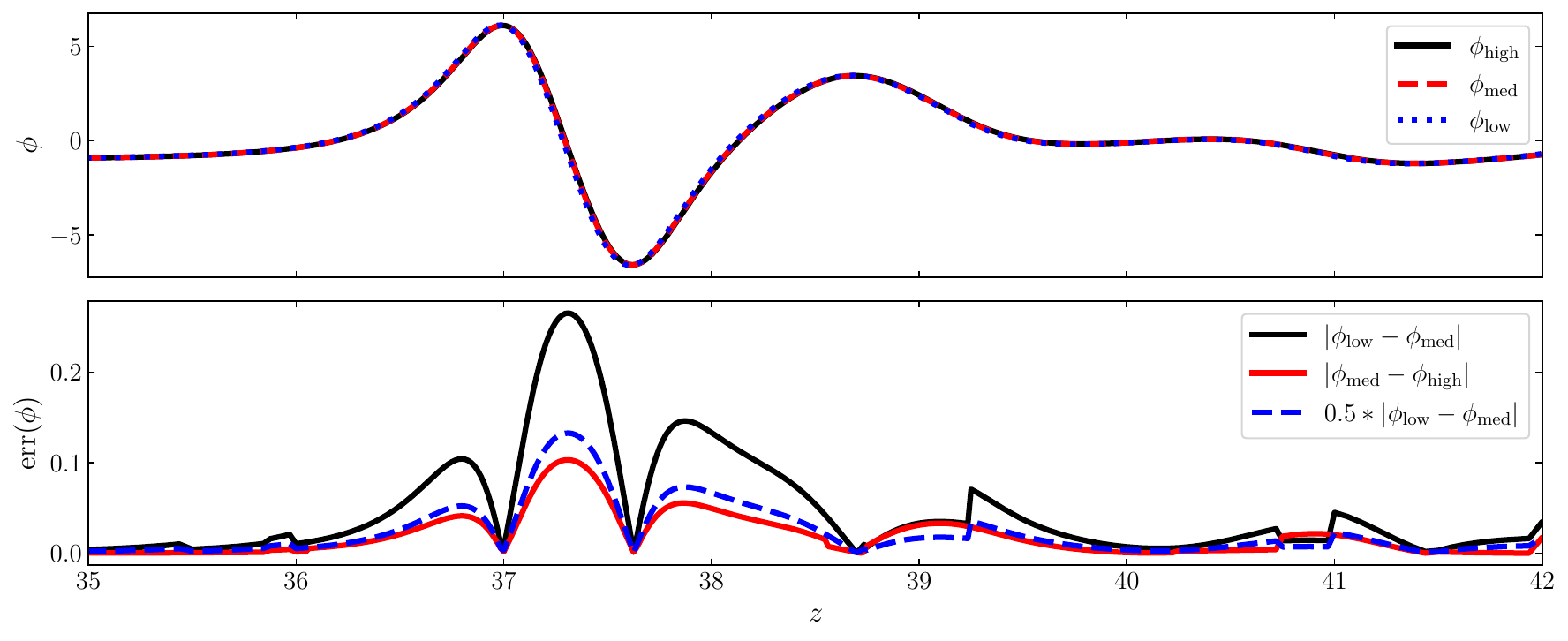}
    \caption{Upper: scalar field profile along the rotation axis of the prolate spheroid collapse at $t= 48$ for runs of three different base resolutions. Lower: difference in field values between resolutions, indicating 1st order convergence.}
    \label{fig:convergence}
\end{figure}

At the top of Fig.~\ref{fig:convergence} we show the scalar field profile at $t = 48$ for the three different resolutions. As one can see, there is minimal change in the profile as one alters the resolution. Below this we plot the respective error in the field $\mathrm{err}(\phi) =  |\phi^i - \phi^j|$ for resolutions $i,j$. We observe approximately 1st order convergence i.e. $p =1$, which is adequate for our qualitative study. \\

\end{document}